


\def\today{\ifcase\month\or January\or February\or March\or
April\or May\or June\or July\or August\or September\or
October\or November\or December\fi \space\number\day,
\number\year}

\def\a{\alpha}
\def\b{\beta}

\def\g{\gamma}

\def\p{\phi}

\def\t{\theta}

\def\vp{\varphi}

\def\D{\Delta}

\def\Si{\Sigma}

\let\nn=\noindent

\font\tenbb=msbm10
\font\sevenbb=msbm8
\font\fivebb=msbm5
\newfam\bbfam
\textfont\bbfam=\tenbb \scriptfont\bbfam=\sevenbb
\scriptscriptfont\bbfam=\fivebb
\def\bb{\fam\bbfam}

\def\CC{{\bb C}}

\def\RR{{\bb R}}

\def\ZZ{{\bb Z}}

\def\part{\partial}
\def\Inf{\infty}

\def\op{\oplus}
\def\ot{\otimes}

\def\bu{\bullet}

\def\ts{\times}

\def\sbs{\subset}

\def\ra{\rightarrow}

\def\hra{\hookrightarrow}

\def\lbc{\lbrace}
\def\rbc{\rbrace}
\def\lbk{\lbrack}
\def\rbk{\rbrack}

\def\oo{\overline}
\def\ww{\widetilde}

\def\ds{\displaystyle }

\def\Da{{\cal D}}

\def\Ha{{\cal H}}

\def\and{\mathop{\rm and}\nolimits}

\def\all{\mathop{\rm all}\nolimits}

\def\Diff{\mathop{\rm Diff}\nolimits}

\def\for{\mathop{\rm for}\nolimits}

\def\grad{\mathop{\rm grad}\nolimits}

\def\Hol{\mathop{\rm Hol}\nolimits}

\def\Im{\mathop{\rm Im}\nolimits}

\def\Maps{\mathop{\rm Maps}\nolimits}

\def\or{\mathop{\rm or}\nolimits}

\def\space{\mathop{\rm space}\nolimits}

\def\sgn{\mathop{\rm sgn}\nolimits}

\catcode`\@=11
\def\displaylinesno #1{\displ@y\halign{
\hbox to\displaywidth{$\@lign\hfil\displaystyle##\hfil$}&
\llap{$##$}\crcr#1\crcr}}

\def\ldisplaylinesno #1{\displ@y\halign{
\hbox to\displaywidth{$\@lign\hfil\displaystyle##\hfil$}&
\kern-\displaywidth\rlap{$##$}
\tabskip\displaywidth\crcr#1\crcr}}
\catcode`\@=12

\def\buildrel#1\over#2{\mathrel{
\mathop{\kern 0pt#2}\limits^{#1}}}

\def\build#1_#2^#3{\mathrel{
\mathop{\kern 0pt#1}\limits_{#2}^{#3}}}

\def\hfl#1#2{\smash{\mathop{\hbox to 6mm{\rightarrowfill}}
\limits^{\scriptstyle#1}_{\scriptstyle#2}}}

\def\hfll#1#2{\smash{\mathop{\hbox to 6mm{\leftarrowfill}}
\limits^{\scriptstyle#1}_{\scriptstyle#2}}}

\def\up#1{\raise 1ex\hbox{\sevenrm#1}}

\def\xx{\vrule height 0.7em depth 0.2em width 0.5 em}

\def\per{|\!\raise -4pt\hbox{$-$}}
 \def\cqfd{\unskip\kern
6pt\penalty 500 \raise
-2pt\hbox{\vrule\vbox to10pt{\hrule
width 4pt \vfill\hrule}\vrule}\par}

\def\trait{\hbox to 12mm{\hrulefill}}
\def\2{{\mathop{\rm
I }\nolimits}\!{\mathop{\rm
I}\nolimits}}

\def\1{{\mathop{\rm I }\nolimits}}

\def\og{\leavevmode\raise.3ex\hbox{\$
scriptscriptstyle\langle\!\langle$}}
\def\fg{\leavevmode\raise.3ex\hbox{\$
scriptscriptstyle \,\rangle\!\rangle$}}

\def\picture #1 by #2 (#3)
{\vcenter{\vskip #2
\special{picture #3}
\hrule width #1 height 0pt depth 0pt
\vfil}}

\def\scaledpicture #1 by #2 (#3 scaled #4){{
\dimen0=#1 \dimen1=#2
\divide\dimen0 by 1000 \multiply\dimen0 by #4
\divide\dimen1 by 1000 \multiply\dimen1 by #4
\picture \dimen0 by \dimen1 (#3 scaled #4)}}

\def\[{{[\mkern-3mu [}}
\def\]{{]\mkern-3mu ]}}

\def\cc#1{\hfill\kern .7em#1\kern .7em\hfill}

\def\TeX{T\kern-.1667em\lower.5ex\hbox{E}\kern-.125em X}

\def\ins{{ \raise -2mm\hbox{$<$}
\atop \raise 2mm\hbox{$\sim$}J}}

\def\sus{{ \raise -2mm\hbox{$>$}
\atop \raise 2mm\hbox{$\sim$}J}}

\magnification=1200
\overfullrule=0mm
\baselineskip=14pt

\vglue 0.5cm
\centerline{\bf $DIFF(S^1)$, TEICHM\"ULLER SPACE AND PERIOD MATRICES:}
\centerline {\bf CANONICAL MAPPINGS VIA STRING THEORY}

\medskip
\centerline {by}
\medskip
\centerline{{\bf Subhashis Nag}}
\medskip
\nn
{\bf 1.Introduction}
\medskip
\nn
{\bf 1a.$Diff(S^1)$ and the Universal Teichm\"uller space:}
The quantizable coadjoint orbits of the
infinite dimensional Frechet Lie group
$Diff(S^1)$ are uniquely determinable as the following
two remarkable homogeneous spaces:
$N=Diff(S^1)/Rot(S^1)$ and $M=Diff(S^1)/Mobius(S^1)$; these symplectic
manifolds are fundamental in the loop-space approach
to the theory of the closed string. Basic references are
the papers of Bowick, Rajeev, Zumino et. al. (see the references
listed in [NV]), and Witten [W].
Actually $N$ and $M$ are homogeneous complex analytic (Frechet)
manifolds with a K\"ahler form that is the symplectic form
of Kirillov-Kostant-Souriau on these
coadjoint orbits. Therefore $N$ and $M$ (the former is
simply a holomorphic fiber space over the
latter with fiber a unit disc) play the role of
{\it phase spaces} for string theory. From (Ricci) curvature computations
on these orbit manifolds with respect to the
above canonical K\"ahler metric, [this was necessary
in order to implement the fundamental requirement of
reparametrization-invariance for closed string theory]
there appears a proof that the bosonic string must be propagating
in spacetime of 26 dimensions (and the fermionic string in 10).

On the other hand, in the Polyakov path integral approach to
the closed bosonic string, the basic partition function has to
be computed as an integral over the space of all world-sheets swept
out by propagating strings. Since the Polyakov action has conformal invariance,
the functional integral one has to perform reduces to one over the
moduli space of all conformal (=complex) structures on
the world-sheets. Thus the Teichm\"uller spaces parametrizing
all the complex structures on a given topological surface
become fundamental in the Polyakov path integral approach to string
theory. The 26 dimensionality of bosonic spacetime from was discovered
first (by Polyakov) from
this approach as a consequence of demanding that the string
partition function integral should be well-defined independent of
how one parametrises the Teichm\"uller space as a space of conformally
distinct Riemannian metrics on the world-sheet.
In a non-perturbative string scattering amplitude calculation
one would be required to do the Polyakov integral
over some "universal parameter space" of Riemann surfaces
that would contain the Teichm\"uller spaces of all topological
types simultaneously.

Now, the Universal Teichm\"uller Space $T(1)$, is such a
universal parameter space for all Riemann surfaces.
$T(1)$ is a (non-separable) complex Banach manifold that may be defined as
the (holomorphically) homogeneous space
$QS \left({  S^1}\right)/$M\"ob $\left({  S^1}\right)$.
The Universal Teichm\"uller space is the universal {\it ambient} space for
the Teichm\"uller spaces of arbitrary Fuchsian groups -- thus it contains
as properly embedded complex submanifolds (multiple copies of) each
of the classical Teichm\"uller spaces of arbitrary Riemann surfaces.
Here $QS \left({  S^1}\right)$ denotes the group of all
{\it quasisymmetric homeomorphisms} of
the unit circle $S^1$, and
M\"ob$\left({  S^1}\right)$ is the three-parameter subgroup of M\"obius
transformations of the unit disc (restricted to its
boundary circle $S^1$).

In attempting to understand the relationship between the
string-reparametrization complex manifolds of the loop space approach
and the Teichm\"uller spaces arising in the sum-over-histories appraoch,
we were able to uncover some new structures
in the Universal Teichm\"uller space which had hitherto not been studied.
Indeed, in [NV] [N2], we showed that the ''phase space"
$M = \Diff \left({  S^1}\right)/$ M\"ob $\left({ S^1}\right)$
sits embedded naturally in $T(1)$ (since any smooth diffeomorphism
is quasisymmetric) -- and that this embedding is an {\it equivariant,
injective holomorphic and K\"ahler isometric immersion}. $M$ can be
interpreted as the submanifold of ''smooth points" of the
Universal Teichm\"uller space; in fact, if $T(1)$ is considered as the
space of all (M\'obius classes of) oriented quasicircles on the Riemann
sphere, then $M$ corresponds precisely to the $C^{\Inf}$ smooth
ones. One thus obtains a picture of the Universal Teichm\"uller space as
foliated by copies of $M$ and its translates under the universal modular
group -- each leaf being a complex analytic Frechet submanifold carrying
a well-defined K\"ahler structure induced from the universal
Weil-Petersson pairing on $T(1)$.

The K\"ahler structure on $T(1)$ given by the universal form of
the Weil-Petersson pairing (see [NV] Part II) is
formal. That pairing converges on precisely the Sobolev class $H^{3/2}$
vector fields on the circle  -- whereas the general tangent vector
to $T(1)$ is represented by vector fields on $S^1$ of the Zygmund
class $\Lambda$. On the smooth vector fields we showed that this pairing
pulled back under the injection of $M$ in $T(1)$ to the Kirillov-Kostant
pairing of $M$.

The complex analytic structure of Teichm\"uller space is canonical
and can be seen as arising from many points of view (see [N4]) -- it is
induced from the complex structure of the ''$\mu$-space'' of Beltrami
coefficients, which comprise a ball in a complex Banach space.
On the other hand, the complex
structure, dictated by representation theory, on the coadjoint ''phase
spaces'' $M$ and $N$ is obtained at the tangent space level by
conjugation of Fourier series (namely ''Hilbert transform'' on functions
on $S^1$). [Recall that tangent vectors to $M$ or $N$ can be thought
of as certain vector fields on the circle. The almost complex structure sends
a vector field to its Hilbert transform!] That the injection of $M$
into $T(1)$ is a holomorphic immersion amount to proving that Hilbert
transform on vector field corresponds at the Beltrami level to
mapping $\mu$ to $i\mu$. That was shown to be true in [NV] Part I.

The convergence/divergence analysis of the universal Weil-Petersson
form gave us moreover a proof that the classical Teichm\"uller spaces
of infinite order Fuchsian groups $G$ always intersect {\it transversely}
the submanifold $M$ of the ''smooth points" of $T(1)$. That assertion was
an infinitesimal version of the {\it Mostow rigidity theorem on the line}.
In fact, as is well known from papers of Bowen, Sullivan et. al., the
quasicircles that are the limit sets of non-trivial quasi-Fuchsian
deformations of $G$ are in general expected to be very {\it non-smooth}
(indeed ''fractal") -- and consequently the quasicircles that
represent points of $T(G)$ cannot be in the submanifold $M$ which
comprises the smooth ones! In this article we shall explain how
to create a {\it Teichm\"uller space $T(H_{\Inf})$ that appears
embedded in $T(1)$ (again in multiple copies) as the natural
completion of the union of copies of the Teichm\"uller spaces of
closed Riemann surfaces of all genera}. Once again we will see that
$T(H_{\Inf})$ is embedded transverse to $M=Diff(S^1)/Mob(S^1)$.
There also appears a genus-independent form of the Weil-Petersson
pairing on $T(H_{\Inf})$ that induces the classical pairing on
the Teichm\"uller spaces of finite genus Riemann surfaces.
This latter material is rather new and appears in
our paper [NS] joint with Dennis Sullivan.

To complete our introductory description
of the structure of the universal Teichm\"uller
space that has arisen from the above considerations we should mention that
the submanifold $M$ is not even locally closed in the Banach manifold
$T(1)$. The {\it closure of $M$} in the universal Teichm\"uller space
is however identifiable (see [Re], [GS]), and
turns out to be the complex Banach submanifold $M^{cl}$ which is
$Symmetric-Homeomorphisms(S^1)/Mob(S^1)$. Symmetric homeomorphisms
are those that have quasiconformal extensions to the unit disc that
are {\it asymptotically conformal} as one approaches the boundary circle.
These are exactly the limits of smooth diffeomorphisms of $S^1$ in the
Teichm\"uller (quasiconformal or ''$\mu$'') topology. Thus $M^{cl}$
is a Banach thickening of the complex Frechet submanifold $M$, and,
of course, copies of $M^{cl}$ will also foliate $T(1)$ by holomorphic
leaves under translation of $M^{cl}$ by the biholomorphic
(right-translation) action of the universal modular group on $T(1)$.
Unfortunately, however, the universal Weil-Petersson does not conversge on
all tangents to the closure of $M$ [indeed, the tangent space to
$M^{cl}$ can be identified as the vector fields on the circle of the
small $\lambda$ Zygmund class -- and these are more general than the
convergence class $H^{3/2}$]. We have in summary:

\nn
{\it Tangent space at origin of:}

\nn
$M=Diff(S^1)/Mobius$ is $C^{\Inf}$ vector fields on
$S^1$ [modulo $sl(2,\RR)$]

\nn
$M^{cl}=Symm(S^1)/Mobius$ is Zygmund $\lambda$ class vector fields on
$S^1$ [modulo $sl(2,\RR)$]

\nn
$T(1)=QS(S^1)/Mobius$ is Zygmund $\Lambda$ class vector fields on
$S^1$ [modulo $sl(2,\RR)$]

\nn
and the Kirillov-Kostant = Universal Weil-Petersson pairing
converges on the
Sobolev class $H^{3/2}$ vector fields on $S^1$ [modulo $sl(2,\RR)$].

We produce below a schematic diagram of the Universal Teichm\"uller
Space incorporating some of its structure discussed above:

\bigskip
\bigskip
\bigskip
\bigskip
\bigskip
\bigskip
\bigskip
\centerline{\bf Figure: Submanifolds in Universal Teichm\"uller Space}

\nn
{\bf 1b.Teichm\"uller Space and Period Matrices:}
In subsequent work ([N1] [N2]) we showed
that {\it infinite dimensional ''period matrices''
can be naturally associated to
the smooth points $M$ of $T(1)$} -- matrices which generalise
exactly the usual $g\times g$ matrices of period integrals
(of Abel-Jacobi) associated to
closed Riemann surfaces of genus $g$.

We were motivated by Segal's construction [S] of a certain representation
of $Diff(S^1)$ as infinite dimensional symplectomorphisms. One realizes
-- by invoking considerations that go away at a tangent
from Segal's representation theoretic aims --
that one is led immediately to a natural universal period mapping.
We will describe this in detail below. In [NS] we succeeded in completing
the theory of the period mapping from the submanifold $M$ to the entire
space $T(1)$.

Before launching off upon the description of this mathematics, it may be
mentioned that such a consideration is again natural from string physics.
In fact, one may wish to compute string scattering amplitude
integrals over spaces of Riemann surfaces by
transferring them to integrals over the space of their Jacobi varieties
-- namely integrate over the Schottky locus in the Siegel space of
principally polarized Abelian varieties instead of on the moduli space of
Riemann surfaces. As is well-known (Torelli's
theorem), that classical association of period
matrices (or, equivalently, Jacobi
varieties) to Riemann surfaces is a one-to-one mapping -- so that
one should lose nothing by such a passage. There were speculations in
the string physics literature on this matter. {\it Our period mapping now
shows that there is a genus-independent non-perturbative way to carry
out this transfer from the space of all Riemann surfaces to a space of
associated Jacobians, and that the rough edges of the finite-dimensional
theory get smoothened out in passing to the universal parameter space since
the universal period map is actually an isometry with respect to the
canonical K\"ahler metrics.}

\nn
{\bf Remark:}
Interestingly then, whether carrying out a string non-perturbative
functional integral over the coadjoint orbit $M$, or over the corresponding
submanifold in Universal Teichm\"uller space $T(1)$, or over the image
(i.e., ''Schottky locus'') of $\Pi$ in Siegel space {\it should produce
the same results} in view of the holomorphy and isometry results
established in [NV] [N1] [N2] [NS] -- and surveyed
above. The infinite-dimensional
differential geometry of the three types of complex analytic parameter
spaces are therefore canonically related to each other!

As mentioned, basic to our construction is the faithful representation
([S]) of $Diff({S^1})$ on the Frechet space
$$
V = C^\Inf \Maps \left({ S^1, \RR }\right)/\RR
({\mathop{\rm  \  the \ constant \  maps \ }\nolimits})
\eqno(1)
$$

\nn
Here $Diff(S^1)$  acts by pullback on
the functions in $V$ as a group
of toplinear automorphisms preserving a basic
symplectic form $S$ that $V$ carries. $V$ can be interpreted as a
smooth version of the Hodge theoretic first cohomology space of the
unit disc, and then $S$ becomes identified with the cup-product
pairing on cohomology. {\it Among the diffeomorphisms one notices that
only the three-parameter M\"obius group acts unitarily on} $V$.
Hence one gets a natural map:
$$
\Pi:M \ra Sp(V)/Unitary
$$
The target space is a complex Banach manifold that is the infinite
dimensional version of the Siegel space of period matrices built by
Segal in [S].
This is the mapping we interpreted as a {\it ''period mapping''} that keeps
track of the varying space of holomorphic 1-forms on a Riemann surface
(as a point of a Grassmann manifold) as the
complex structure on the surface is varied; that is P. Griffiths way
of describing just what the classical period mappings do. The fundamental
naturality of the map was evident from the result we proved that, like the
injection of $M$ in $T(1)$, this mapping
{\it $\Pi$ is an equivariant, injective (genus-independent Torelli
theorem!), holomorphic and K\"ahler isometric immersion of the
smooth-points-submanifold of the Universal Teichm\"uller space to
an infinite dimensional version of the Siegel space of period matrices.}

\nn
{\bf Note:} The complex structure of the Siegel space arises, as in the
finite dimensional theory, by interpreting the points of the Siegel space
as positive polarizing subspaces in the complexification of a symplectic
vector space (the first cohomology vector space of the reference Riemann
surface). Clearly then, the Siegel space appears embedded in
a complex Grassmannian -- hence the complex structure is induced from that
of the Grassmannian. As for the metric, the finite dimensional Siegel space

$$
{\cal S}_{g}=Sp(2g, R)/Maximal~compact~subgroup~U(g)
$$

\nn
carries a unique (up to scaling constant) homogeneous
K\"ahler metric for which the full
group $Sp(2g, R)$ of holomorphic automorphisms acts as isometries.
That metric simply goes over to the infinite
dimensional Siegel space
 -- and that canonical metric, we showed in [N2], pulls back via $\Pi$ to
universal Weil-Petersson = Kirillov-Kostant on the space $M$.

As we promised, we will also describe more recent joint work
with Dennis Sullivan ([NS]) where {\it we have succeeded in finding the
natural extension of the above period mapping to the
entire Universal Teichm\"uller space by utilising the Sobolev (Hilbert)
space $H^{1/2}$ on the circle to complete the $C^{\Inf}$ space $V$.}
It is completely clear that in
order to be able to extend the infinite dimensional
period map to the full space $T(1)$, it is necessary to
replace $V$ by a suitable ``completed'' space
that is invariant under quasisymmetric pullbacks.
Moreover, the quasisymmetric
homeomorphisms should continue to act as bounded
symplectic automorphisms of this extended space.
The Hilbert space $ H^{1/2}$ = $\Ha$, (say), appears indicated from
(at least!) two points of view and fits the bill exactly.

{\it First, the above Hilbert space, which turns out
to be exactly the completion
of the pre-Hilbert space $V$}, actually {\it characterizes}
quasisymmetric (q.s).  homeomorphisms (amongst all
homeomorphisms of the circle) in the sense
that q.s. homoeomorphisms, and only
those, act as bounded operators on $\Ha$ by pull-back (namely
pre-composition). That will obviously be important for our understanding
of the universal period mapping.
Harmonic analysis tells us that a general $H^{1/2}$ function
on the circle is defined off a set of (logarithmic) capacity zero.
Notice that the fact that capacity zero sets are preserved by
quasisymmetric transformations --  whereas merely being
measure zero is not a q.s.-invariant
notion --  goes to exemplify how deeply quasisymmetry
is connected to the properties of $\Ha$.

{\it Furthermore, the space $H^{1/2}$ can be canonically identified as
the Hodge theoretic first cohomology space of
the universal Riemann surface (the unit disc)} -- namely the above
Hilbert space {\it is} the space of square-integrable
real harmonic 1-forms for the disc.
The basic reason again comes from the ''trace theorems''
of harmonic analysis --  the $H^{1/2}$ functions are exactly the
non-tangential boundary values (''traces'') of harmonic functions of
finite Dirichlet energy on the unit disc.
The symplectic=cup-product  structure, $S$, (seen on $V$)
extends to $\Ha$ and is preserved by the pullback action of
quasisymmetric homeomorphisms. Therefore, the extension
of $\Pi$ from $M$ to all of $T(1)$
utilizing this completion $\Ha$ of $V$ can again be interpreted
as being a period mapping in P. Griffiths' sense.
Not only does the cup-product get identified with
the canonical symplectic form carried by $\Ha$,
one finds that the fiducial complex structure on $\Ha$
obtained from the Hodge star on harmonic 1-forms is given precisely by the
Hilbert transform (which again preserves the space $H^{1/2}$).
The target space for the period map is the universal Siegel
space of period matrices; that is the space of
all the complex structures $J$ on $\Ha$ that are compatible with
the canonical symplectic structure (or, alternatively, all positive
polarizing subspaces in the complexification of this first cohomology
Hilbert space). The various incarnations of the Siegel space will be
spelled out below (Section 8).

In effect, an arbitrary point $X$ of $T(1)$ produces a decomposition
$$
Complexified~H^{1/2}= H^{1,0}(X) \op H^{0,1}(X)
$$
\nn
and the universal period map associates to $X$ the positive polarizing
subspace $H^{1,0}(X)$ as a point of the universal Siegel space.
That is why $\Pi$ is the universal period mapping.
$\Pi$ {\it thus provides us with a new realization of the Universal
Teichm\"uller space as a complex submanifold of the Universal Siegel
space.}

To establish the complete naturality of the construction of $\Pi$, we
have also proved in [NS] that the pairing $S$ is
the {\it unique} symplectic structure available which
is invariant under even the tiny finite-dimensional subgroup
M\"ob$\left({  S^1}\right)$ ($\sbs QS\left({S^1}\right)$).
That proof uses the discrete series representations of $SL(2,\RR)$
and a version of Schur's lemma.

\nn
{\bf 1c.Universal period mapping and Quantum calculus:}
One must remember first of all that the generic
quasisymmetric homeomorphisms of the circle that arise in the
Teichm\"uller theory of Riemann surfaces as boundary values of
quasiconfomal homeomorphisms of the disk have fractal graphs in
general, and are consequently not so amenable to usual
analytical or calculus procedures. Therefore we made use of
the remarkable fact this group $QS(S^{1})$ acts by substitution
(i.e., pre-composition) as a family of
bounded symplectic operators on the Hilbert space $\Ha$=``$H^{1/2}$''

But Alain Connes' has suggested a {\it ''quantum differential''}
$d^{Q}_{J}f = [J,f]$ -- commutator of the multiplication operator with
the complex structure operator (Hilbert transform) -- to replace the
ordinary (often non-existent) classical derivative for a function $f$
(say on the circle). The idea is that operator theoretic properties of
the quantum differential will capture smoothness characteristics of $f$.
Utilizing this idea,
we obtain ([N3]) in lieu of the problematical
classical calculus a quantum calculus for quasisymmetric homeomorphisms.
Namely, one has operators $\{h,L\}$, $d\circ\{h,L\}$, $d\circ\{h,J\}$,
corresponding to the non-linear classical objects
$log (h^\prime)$, ${h'' \over {h'}}dx$, ${1\over 6}Schwarzian(h)dx^{2}$.
The point is that these quantum operators make sense whenever $h$ is merely
quasisymmetric, whereas the classical forms could be
defined only when $h$ was appropriately smooth. Any one of these objects is a
quantum measure of the conformal distortion of $h$ in analogy with the
classical calculus' Beltrami coefficient $\mu$ for a
quasiconformal homeomorphism of the disk. Here $L$ is the smoothing
operator on the line (or the circle) with kernel $log \vert x-y \vert$,
$J$ is the Hilbert transform (which is $d \circ L$ or $L \circ d$), and
$\{h,A\}$ means $A$ conjugated by $h$ minus $A$.

The period mapping and the quantum calculus are related in several ways.
For example, $f$ belongs to $\Ha$ if and only if the quantum
differential is Hilbert-Schmidt. Also, the Schottky locus or image of
the period mapping is describable  by a {\it quantum integrability condition}
$[d^{Q}_{J},J]=0$ for the complex structure $J$ on $\Ha$.

We will present below a section
where we discuss quantum calculus.
the idea being firstly to demonstrate that
the $H^{1/2}$ functions have such an interpretation. That then allows us
to interpret the universal Siegel space that is the target space for the period
mapping as "almost complex structures on the line" and the Teichm\"uller
points (i.e., the Schottky locus ) can be interpreted as comprising
precisely the subfamily of those complex structures that are {\it
integrable}.

\nn
{\bf 1d.Universal Period Mapping restricted to the Teichm\"uller
space of the universal compact Riemann surface:}
As mentioned before, inside the Universal Teichm\"uller space
there resides the separable complex
submanifold $T(H_\Inf)$ -- the Teichm\"uller space of the universal
hyperbolic lamination -- that is exactly the closure of the union of
certain copies of all the genus $g$
classical Teichm\"uller spaces of closed Riemann surfaces in $T(1)$.
Genus-independent constructions like the universal period mapping proceed
naturally to live on this completed version
of the classical Teichm\"uller spaces.

The polarizing subspace determined by $\Pi$ for a point of $T_{g}$
has an intimate relationship with the subspace of $L^{2}(S^1)$
determined by the Krichever map on data living on this compact Riemann
surface. We will spell this out.
In fact, the restriction of the universal period
mapping $\Pi$ to the submanifold $T(H_\Inf)$ of
$T(1)$ should have deep properties that relate to the
classical period mappings $\pi_{g}$ which live on the Teichm\"uller
spaces of genus $g$ surfaces (which appear embedded within $T(H_\Inf)$).
To start with, we showed in [N3] that
$T(H_\Inf)$ carries a natural convergent Weil-Petersson pairing.
We hope that in future publications we will show how the understanding
of the universal Schottky locus that we will describe can be used to
determine the Schottky locus in the classical genus $g$ situation.

\medskip
\nn
{\bf Acknowledgements:}
It is my pleasant duty to acknowledge gratefully many
stimulating conversations with a host of
mathematicians who took an active interest in
this work over the past few years and listened kindly to my talks
on the intimate relationships between these ''universal parameter
spaces''. I would like to thank all of them -- especially the
many interested participants in the Mathematical Society of Japan
International Research Institute on the Topology of the Moduli Space
of Curves (RIMS, Kyoto, 1993). It goes without saying --
and yet it must needs
be said -- that I would like to thank Alberto Verjovsky and Dennis
Sullivan for collaborating with me at various times on the work
that I am reporting upon here.

I want to express my gratitude to the organisers of the International
Conference on Complex Analysis and the VIIth Romanian-Finnish Seminar
in Complex Analysis (Romania 1993) for inviting this survey paper.
Finally I must thank the IHES (Paris) as well as
the ICTP (Trieste) and the CUNY Graduate Center (New York), for
their excellent hospitality during the Autumn/Winter of 1992, where
many of the more recent results on the universal period mapping were obtained.

\medskip
\nn
{\bf 2.Teichm\"uller Theory and Universal Weil-Petersson: }

\noindent
{\bf Basic definitions:}
The universal Teichm\"uller space $T(1)$ is a holomorphically
homogeneous complex Banach manifold that serves
as the universal ambient space
where all the Teichm\" uller spaces (of arbitrary Fuchsian groups) lie
holomorphically embedded.
Let $ \Delta $ denote the open unit disc, and $ S^{1} = \partial \Delta $.
Two classic models of the Universal Teichm\" uller Space
$T(1) = T(\Delta)$ will be described below:

\noindent
(a) the ''real-analytic model'' containing all (M\"obius-normalised)
quasisymmetric homeomorphisms of the unit circle $S^1$ ;

\noindent
(b) the ''complex-analytic model'' comprising all normalized schlicht
(Riemann mapping) functions on the exterior of the disc
which allow quasiconformal extension to the whole
Riemann Sphere -- i.e., which conformally map the exterior of
the disc to (the exterior of) a quasidisc.

The Teichm\"uller space of any Fuchsian group $G$ comprises those
(M\"obius-normalized) quasidiscs within which a quasi-Fuchsian deformation
of $G$ acts discontinuously. Alternatively, (picture (a)), $T(G)$
is the family of quasisymmetric homeomorphisms of $S^1$ that are compatible
with $G$ (namely conjugate $G$ again into groups of M\"obius transformations).

The connection between them is via the rather mysterious operation
called ``Conformal Welding" (see [KNS],[N]).
That means that the quasidisc in model (b) that is the image of the
given schlicht function determines a
quasisymmetric homeomorphism of the circle
by comparing the boundary values of the Riemann mappings to its interior
and exterior.

To recall matters,  we set the stage by introducing
the chief actor -- namely the
space of (proper) Beltrami coefficients
$L^{\infty}(\Delta)_{1} $ ; it is the
open unit ball in the complex Banach space of $L^{\infty}$ functions on the
unit disc $\Delta$.  The principal construction is to solve the Beltrami
equation.
$$
w_{\oo{z}}  =  \mu w_{z} \eqno(Bel)
$$

\noindent for any $\mu \in L^{\infty} (\Delta)_{1}$.  The two
above-mentioned models of Teichm\" uller space correspond to discussing two
pertinent solutions for (Bel) :

\noindent {\bf (a)} $w_{\mu}$ - theory : The quasiconformal hemeomorphism of
{{\bf C}} which is $\mu$-conformal (i.e. solves (1)) in $\Delta$, fixes $\pm 1$
and $-i$, and keeps $\Delta$ and $\Delta^{\star}$ (= exterior of $\Delta$) both
invariant.  This $w_{\mu}$ is obtained by applying the existence and uniqueness
theorem of Ahlfors-Bers (for (Bel)) to the
Beltrami coefficient which is $\mu$ on
$\Delta$ and extended to $\Delta^{\star}$ by reflection $(\tilde{\mu}
(1/\oo{z}) = \overline{\mu(z)} z^{2}/\oo{z}^{2}$ for $z \in \Delta)$.

\nn
[{\bf Note:} The extension of $\mu$ to the whole plane by reflection above is
{\it not} complex-linear, -- that is why model (a) is being called a
''real-analytic'' model of Teichm\"uller space.
In this model the (almost-)complex
structure appears mysteriously via the Hilbert transform on vector fields.]

\noindent {\bf (b)} $w^{\mu}$ - theory : The quasiconformal homeomorphism on
{\bf C}, fixing $0,1,\infty$, which is $\mu$-conformal on $\Delta$ and
conformal on $\Delta^{\star}$. $w^{\mu}$ is obtained by applying the
Ahlfors-Bers theorem to the Beltrami coefficient which is $\mu$ on $\Delta$ and
zero on $\Delta^{\star}$.

The fact is that $w_{\mu}$ depends only real analytically on $\mu$,
whereas $w^{\mu}$ depends complex-analytically on $\mu$.  We therefore obtain
two standard models ({\bf{(a)}} and {\bf{(b)}} below) of the universal
Teichm\" uller space, $T(\Delta)$ as stated above.

Thus, we define the universal Teichm\" uller space :
$$
T(1) = T(\Delta) = {L}^{\infty} (\Delta)_{1} / \sim .
$$

\noindent Here $\mu \sim \nu$ if and only if $w_{\mu} = w_{\mu}$ on $\partial
\Delta = S^{1}$, and that happens if and only if the conformal mappings
$w^{\mu}$ and $w^{\nu}$ coincide on $\Delta^{\star} \cup S^{1}$.
We let
$$
\Phi : L^{\infty}(\Delta)_{1} \rightarrow T(\Delta)
$$

\noindent denote the quotient (``Bers") projection.  $T(\Delta)$ inherits its
canonical structure as a {\bf complex Banach manifold}
from the complex structure of
$L^\infty(\Delta)_{1}$ ; namely, $\Phi$ becomes a {\it{holomorphic
submersion}}.

The derivative of $\Phi$ at $\mu = 0$ :
$$
d_{0} \Phi : L^{\Inf}(\Delta) \rightarrow T_{O} T(\Delta)
$$

\noindent is a complex-linear surjection whose kernel is the space $N$ of
``infinitesimally trivial Beltrami coefficients''.
$$
N = \{ \mu \in L^{\Inf}(\Delta) : \int_{}^{}\int_{\Delta}^{}\mu \phi = 0~~
for~~ all~~\phi \in A(\Delta) \}
$$
\noindent
where $A(\Delta)$ is the Banach space of integrable $(L^1)$
holomorphic functions on the disc.  Thus, the tangent space at the origin
of $T(\Delta)$ is canonically $L^{\Inf}(\Delta)/N$.
See [A3], [Le] and [N4] for this material and for what follows.

It is now clear that to $\mu \in L^{\Inf}(\Delta)_{1}$ we can
associate the {\it quasisymmetric homeomorphism}
$$
f_{\mu} = w_{\mu}
$$
\nn
{\it on the boundary circle of the hyperbolic disc
as representing the Teichm\" uller point} $[\mu]$ in version
{\bf{(a)}} of $T(\Delta)$.  Indeed $T(\Delta)_{(a)}$ is the homogeneous space
comprising M\"obius classes of quasisymmetric homeomorphisms of the
unit circle. Alternatively, it is identified with the
group (``partial topological group'') of {\it
quasisymmetric homeomorphisms fixing} $1$, $-1$ {\it and} $-i$.

In the {\bf (b)} model, $[\mu]$ {\it is represented by the univalent function}
$$
f^\mu = w^\mu
$$
\noindent
on $\Delta^{\star}$. That is version {\bf (b)} of $T(\Delta)$.
Smooth diffeomorphisms (recall $M$) in model (a) correspond to (namely
they  are the weldings for) univalent functions that map
$\Delta^{\star}$ onto smooth Jordan regions.

It is worth remarking here that the criteria that a
power series expansion
represents an univalent function, and that it allows quasiconformal
extension, can be written down solely in terms of the coefficients $c_{k}$,
(using the Grunsky inequalities etc.). Thus
$T(\Delta)_{(b)}$ can be thought of as a certain space of sequences
$(c_{2},c_{3},...)$, (representing the power series
coefficients of Riemann mappings to quasidisc regions). The
tangent space can be given a concomitant description (see [N3]).

\noindent
{\bf Universal Weil-Petersson:}
As usual, let Diff$(S^1)$ denote the infinite dimensional Lie group of
orientation preserving $C^\infty$ diffeomorphisms of $S^1$.  The complex-
analytic homogeneous ''phase space'' of string theory is, as we explained:
$$
M =  Diff(S^1)/Mob(S^1)
$$

$M$ injects holomorphically into $T(\Delta)_{(a)}$.  This was proved in
[NV, Part I].  The submanifold $M$ comprises the ``smooth points" of
$T(\Delta)$ ; in fact, in version (b), the points from $M$ are those quasidiscs
$F^{\mu}(\Delta^{\star})$ whose boundary curves are $C^\Inf$.

Now, $M$, together with its modular group translates,
foliates $T(\Delta)$ -- and
the fundamental Kirillov-Kostant K\" ahler (sympletic) form exists
on each leaf of the foliation.  Up to an overall scaling
this homogeneous K\" ahler metric gives the following
hermitian pairing $g$ on the tangent space at the origin of $M$ :
$$
g(V,W) = Re \left[ \sum_{k=2}^{\Inf} a_{k} {\oo{b}_{k}}(k^3-k) \right]
$$
\noindent where
$$
V  =  \sum_{2}^{\Inf} a_{k} e^{ik \theta}  +  \sum_{2}^{\infty}
\oo{a}_{k} e^{-ik \theta},
$$
\nn
and similarly $W$ (with Fourier coefficients $b_{k}$),
represent two smooth real vector fields on $S^1$ (tangents to $M$).
[{\bf Note:} The $(k^3 - k)$ appearing in the formula for $g$ is, of course,
closely related to that same term appearing in the Gelfand-Fuks cohomology
theory of the Lie algebra of smooth vector fields on $S^1$.]

The metric $g$ on $M$ was proved by this author [NV, Part II] to
be nothing other than {\it{the
universal Weil-Petersson metric (WP) of universal Teichm\" uller space}}.
Of course, the {\it classical Weil-Petersson} metric
is defined on the Teichm\"uller
spaces of finite dimension by pairing holomorphic quadratic differentials
for a Fuchsian group $G$ using the Petersson pairing of automorphic forms.
The point is that when $G$ is taken to be the universal Fuchsian group
$G=\{1\}$, the pairing becomes the formula written out above.

Let us take a moment to recall the formula for the Weil-Petersson pairing
in terms of Beltrami differentials representing tangent directions
to Teichm\"uller space. Let $\mu$ and $\nu$ be $g$-invariant Beltrami
(-1,1) coefficients on the disc. Then:
$$
WP(\mu,\nu)=\int_{}^{}\int_{\Delta/G}^{}{\times}\int_{}^{}\int_{\Delta}^{}
[\mu(z){\oo {\nu( \zeta)}}]/{(1-z{\oo \zeta})^{4}}
d\zeta \wedge d{\oo \zeta}dz \wedge d{\oo z}
$$

A word about the proof.
Using the perturbation formula for the theory of the Beltrami equation,
the author calculated the Fourier coefficients of the vector fields on
the circle corresponding to the Beltrami directions above. It then
followed that the WP hermitian pairing above, for $G=\{1\}$, coincides
with the Kirillov-Kostant $g$.

Moreover, we showed in [NV] Part II how to
recover the classical W-P pairings on
tangent vectors to finite dimensional Teichm\"uller spaces $T(G)$
by regulating the universal Weil-Petersson formula.
(Universal W-P diverges in general when applied
to such $G$-invariant tangent vectors). That proved
a form of the Mostow rigidity theorem expressing
the transversality of $M$ with the Teichm\"uller spaces $T(G)$, as we
mentioned in the Introduction.

Now, although ''conformal welding'' operation that relates the two
models shewn of $T(1)$ is difficult to analyse, nevertheless, at the
infinitesimal level, the above models have an amazingly simple relationship
that forms the basis for our paper[N3].  Indeed, the $k^{th}$ {\it{Fourier
coefficient}} of the vector field
representing a tangent vector in model (a),
and the (first variation of) the $k^{th}$ {{\it{power series
coefficient}} representing the {\it{same}} tangent vector in model (b),
turn out to be just ($\sqrt{-1}$ times) complex conjugates of each other.
That relationship can also be formulated as a direct identity relating
the vector field on the circle with the holomorphic function in the
exterior of the disc describing the perturbation in the complex-analytic
model. It allows us to express the universal Weil-Petersson metric,
namely the Kirillov-Kostant pairing on $M$, in the following deceptively
simple (and equivalent) form:

We express the pairing for $g = WP$ in terms of 1-parameter
flows of schlicht functions (from [N3]) as follows:
\noindent
Let $F_{t}(\zeta)$ and $G_{t}(\zeta)$ denote two curves
through origin in $T(\Delta)_{(b)}$  representing two tangent
vectors, say $\dot{F}$ and $\dot{G}$.
Then the Weil-Petersson pairing assigns:
$$
WP(\dot{F},\dot{G}) = -\left[\sum_{k=2}^{\infty} \overline{\dot{c}_{k}(0)}
\dot{d}_k(0) (k^3-k) \right],
$$
\noindent
Here $c_k(t)$ and $d_k(t)$ are the power series coefficients for the
schlicht functions $F_t$ and $G_t$ respectively.  The series
above converges precisely  when the corresponding Zygmund class functions
are in the Sobolev class $H^{3/2}$.

\nn
{\bf Holomorphy of the inclusion map $I$ of $M$ into $T(1)$:}
Utilizing the identity proved in [N3]
between Fourier coefficients of Zygmund class vector fields and the
corresponding power series for the Riemann mappings
onto curves of quasidiscs,
we get an immediate proof of the fascinating fact that {\it the
almost complex structure of $T(\Delta)$
transmutes to the operation of Hilbert
transform at the level of Zygmund class vector fields on $S^1$.}
In other words, one sees that
the vector field determined by Beltrami direction $\mu$ is related to
the field obtained from the Beltrami direction
$i\mu$ as a pair of conjugate Fourier series.
A proof of this was important for our previous work, and appeared
in [NV, Part I] -- establishing that $M$ injects {\it holomorphically}
into Universal Teichm\"uller space.

\nn
{\bf Remark:} The result above gives an independent proof of the fact that
conjugation of Fourier series preserves
the Zygmund class $\Lambda(S^1)$.  That was an old theorem of Zygmund [Z].

\medskip
{\it In the next few sections we develop the theory necessary to work out
the universal period mapping $\Pi$ on $T(1)$. We shall explain in detail
the occurence in various incarnations of the Hilbert space $H^{1/2}(S^1)$
and demonstrate its canonical isomorphism with the Hodge-theoretic
first cohomology space of the disc.}

\medskip
\nn
{\bf 3. The Hilbert space $H^{1/2} $ on the circle:}

Let $\D$ denote the open unit disc and $U$ the upper half-plane
in the plane ($\CC$), and $S^1 = \part \D$ be the unit circle.
Intuitively speaking, the real Hilbert space
under concern:

$$\Ha \equiv H^{1/2} \left({  S^1, \RR}\right)
 /\RR \eqno (2)$$

\nn
is the subspace of  $L^2(S^1)$
comprising real functions of mean-value
zero on $S^1$ which have  a half-order derivative also in
$L^2 \left({  S^1}\right)$. Harmonic analysis will tell us that these
functions are actually defined off some set of capacity zero
(i.e., "quasi-everywhere") on the circle, and that they also appear as
the boundary values of real harmonic functions of finite Dirichlet
energy in $\D$. Our first way (of several) to make this precise is
to {\it identify $\Ha$ with the sequence space}

$$\ell_2^{1/2} =
\{ {\mathop{\rm complex \ sequences  }\nolimits}
\ \ u \equiv (u_1, u_2, u_3, \cdots ): \{ \sqrt{n} \ u_n \} \
{\mathop{\rm is \ square \ summable \  }\nolimits} \}.
\eqno (3)
$$

The identification between (2) and (3) is by showing
that the Fourier series

$$f
\left({ e^{i \t } }\right) =
\sum_{ n = - \Inf }^{ \Inf } u_n e^{i n \t} ;
{}~~~u_{- n} = \oo u_n,
\eqno (4)
$$

\nn
converges quasi-everywhere and defines a real function
of the required type.
The norm on $\Ha$ and on $\ell_2^{1/2}$ is, of course,
 the $\ell_2$ norm of
$\{ \sqrt{n } \ u_n \}$, i.e.,

$$
\left\Vert{ f }\right\Vert_{\Ha}^{2} =
\left\Vert{ u }\right\Vert_{\ell_2^{1/2}}^{2} =
2 \sum_{ n = 1}^{ \Inf  } n \left\vert{ u_n }\right\vert_{}^{2}
\eqno (5)
$$

Naturally $\ell_2^{1/2}$ and $\Ha$ are isometrically
isomorphic separable Hilbert
spaces. Note that $\Ha$ is a subspace of
$L^2 \left({  S^1}\right)$ because
$\{ \sqrt{n } \ u_n \}$ in $\ell_2$ implies
$\{  \ u_n \} $ itself is in $\ell_2$.

At the very outset let us note the fundamental fact
that the space $\Ha$ is evidently closed under {\it Hilbert transform }
(``conjugation'' of Fourier series):

$$ (Jf) (e^{i\t}) = - \sum_{ n = - \Inf }^{ \Inf } i
\sgn(n){u_n} e^{in\t}
\eqno (6)$$
\nn
In fact, $J: \Ha \ra \Ha$ is an isometric isomorphism
whose square is the
negative identity, and thus $J$ defines a {\it canonical
complex structure for $\Ha$}.
When identified with the real cohomology space of the disc, this
{\it Hilbert transform will appear as the Hodge star
complex structure.} See below.

\medskip
\nn
{\bf Remark:}
In our earlier articles we had made use of the fact
that the Hilbert transform defines the
almost-complex structure operator for the tangent space
of the coadjoint orbit manifolds ($M$ and $N$), as well
as for the universal Teichm\"uller space $T(1)$.
This fact is closely related to the matter at hand.

\medskip
Whenever convenient we will pass to a description of our Hilbert space
$\Ha$ as functions on the real line, $\bf R$. This is done by simply
using the M\"obius transformation of the circle onto the line that is
the boundary action of the Riemann mapping ("Cayley transform")
of $\D$ onto $U$. We thus get an isometrically isomorphic copy, called
$H^{1/2}(\RR)$, of our Hilbert space $\Ha$ on the circle
defined by taking $f \in \Ha$ to correspond to $g \in
H^{1/2} (\RR)$ where $g = f \circ R,
R(z) = {z - i \over z + i} $ being the Riemann mapping. The Hilbert
transform complex
structure on $\Ha$ in this version is then described by the usual
singular integral operator on the real line with the
"Cauchy kernel" $(x - y)^{-1}$.

\medskip
Fundamental for our set up is the dense subspace
$V$ in $\Ha$ defined by equation (1) in the Introduction.
$V$ will be the ''smooth part'' of the Hodge-theoretic
cohomology of the disc, and it will carry the theory of the period mapping
on the submanifold $M$ of smooth points of $T(1)$.
At the level of Fourier series, $V$ corresponds
to those sequence $\{ u_n \}$ in
$\ell_2^{1/2}$ which go to zero more rapidly than
$n^{- k}$ for any $k >  0$.
Since trigonometric polynomials are in $V$,
it is obvious that $V$ is norm-dense in $\Ha$.

$V$ carries the basic symplectic form that we utilised
crucially in [N1], [N2]:

$$S: V \ts V \ra \RR \eqno (7)$$

\nn
given by

$$S(f, g) =
{1 \over 2 \pi}
\int_{ S^1}^{ } f \cdot dg .
\eqno (8)$$

\nn
This is essentially the signed area of the $(f(e^{i\t}), g(e^{i\t}))$
curve in Euclidean plane.
On Fourier coefficients this bilinear form becomes

$$S(f, g) = 2 \Im\left({ \sum_{n = 1 }^{\Inf} n u_n \oo v_n }\right)
= -i\sum_{n = - \Inf }^{\Inf} n u_n  v_{- n }
\eqno (9)$$
\nn
where $\{ u_n \}$ and $\{ v_n \}$ are respectively
the Fourier ceofficients of the
(real-valued) functions $f$ and $g$, as in (4).  Let us note that the
Cauchy-Schwarz inequality applied to (9) shows that
{\it this non-degenerate
bilinear alternating form extends from $V$ to the full
Hilbert space} $\Ha$.
We will call this extension also
$S: \Ha \ts \Ha \ra \RR$. {\it We shall see that this very symplectic
form is the cup-product in cohomology when we interpret} $\Ha$ {\it as
Hodge-theoretic cohomology space.}

Notice that Cauchy-Schwarz asserts:
$$
\left\vert{ S(f, g) }\right\vert \leq
\left\Vert{ f }\right\Vert
\cdot
\left\Vert{ g }\right\Vert .
\eqno (10)
$$

\nn
Thus $S$ is a jointly continuous, in fact
analytic, map on $\Ha \ts \Ha$.

\medskip
The important interconnection between the inner
product on $\Ha$, the Hilbert-transform complex
structure $J$, and the form $S$ is encapsulated
in the identity:
$$S \left({ f, Jg }\right) =
\left\langle{  f, g }\right\rangle, \
{\mathop{\rm  for \ all \ }\nolimits}
f, g \in \Ha
\eqno (11)$$
\nn
The point is that intertwining the cup-product with the Hodge star
operator as in (11) always produces the inner product in
$L^2$ cohomology of a Riemann surface.

{\it To sum up, therefore, $V$ itself was
naturally a pre-Hilbert space with
respect to the canonical inner product arising from its symplectic form
and its Hilbert-transform complex structure, and we have just
established that the {\bf completion} of $V$
is nothing other than the Hilbert
space $\Ha$. Whereas $V$ carried the $C^{\Inf}$ theory, beacause it was
diffeomorphism invariant, the completed Hilbert space $\Ha$ allows
us to carry through our constructions for the full Universal
Teichm\"uller Space because it indeed is quasisymmetrically
invariant -- as we shall show below.}

\nn
{\bf Complexification of $\Ha$:}
It will be important for us to {\it complexify} our
spaces since we need to deal
with isotropic subspaces and polarizations.
Thus we set

$$\CC \ot V \equiv V_\CC = C^\Inf \
{\mathop{\rm Maps }\nolimits}
\left({  S^1, \CC }\right) / \CC
$$

$$
\CC \ot \Ha \equiv \Ha_\CC = H^{1/2}
\left({  S^1, \CC }\right) / \CC
\eqno (12)
$$

\nn
$\Ha_\CC$ is a complex Hilbert space isomorphic to
$\ell_2^{1/2} (\CC)$ - the latter comprising
the Fourier series

$$f
\left({ e^{i \t } }\right) =
\sum_{n = - \Inf }^{ \Inf }
u_n e^{i n \t} \ \ , \ \ u_0 = 0
\eqno (13)$$

\nn
with $\{ \sqrt{ |n| } \ u_n \}$ being square summable
 over $\ZZ - \{ 0 \}$. Note
that the Hermitian inner product on $\Ha_\CC $
derived from (5) is given by

$$\left\langle{  f, g}\right\rangle =
\sum_{ n = - \Inf}^{ \Inf  } |n| u_n \oo v_n.
\eqno (14)$$

\nn
[This explains why we introduced the factor 2 in the formula
(5).] The fundamental {\it orthogonal decomposition} of
$\Ha_\CC$ is given by

$$\Ha_\CC = W_+ \op W_- \eqno (15)$$

\nn
where

$$W_+ =
\{ f \in \Ha_\CC: \
{\mathop{\rm  all \ negative \ index \ Fourier \
coefficients \ vanish \
}\nolimits} \}$$

\nn
and
$$\oo W_+ = W_- =
\{ f \in \Ha_\CC: \
{\mathop{\rm  all \ positive \ index \ Fourier \
coefficients \ vanish \
}\nolimits}
\}.
$$
We have denoted by bar the complex anti-linear
automorphism of $\Ha_\CC$ given by conjugation of complex scalars.

Let us extend $\CC$-linearly the form $S$
and the operator $J$ to $\Ha_\CC$ (and
consequently also to  $V_\CC$).
The complexified $S$ is still given by the
right-most formula in (9). Notice that
$W_+ $ and $W_-$ can be characterized as
precisely the $- i$
{\it and $+ i$ eigenspaces} (respectively)
 of the $\CC$-linear extension of $J$, the Hilbert transform.
Further, each of $W_+ $ and $W_-$ is
{\it isotropic} for $S$, i.e.,
$S(f, g) = 0$, whenever both $f$ and $g$
 are from either $W_+$ or $W_-$ (see
formula (9)). Moreover, $W_+$ and $W_-$
are {\it positive isotropic} subspaces in the
sense that the following identities hold:

$$
\left\langle{ f_+, g_+ }\right\rangle_{}^{} =
i S
\left({ f_+, \oo g_+ }\right), \ for \
f_+, g_+ \in W_+ \eqno (16)$$

\nn
and

$$
\left\langle{ f_-, g_- }\right\rangle_{}^{} = -
i S
\left({ f_-, \oo g_- }\right), \ for \
f_-, g_- \in W_- . \eqno (17)$$

\nn
{\bf Remark:}
(16) and (17) show that we could have defined the inner product and norm on
$\Ha_\CC$ from the symplectic form $S$, by using these relations to {\it
define } the inner products on $W_+$ and $W_-$, and declaring $W_+$ to be
perpendicular to $W_-$. Thus, for general $f, g \in \Ha_\CC$
one has the fundamental identity

$$
\left\langle{ f, g }\right\rangle_{}^{} = i S
\left({ f_+, \oo g_+ }\right) - i S\left({ f_-, \oo g_- }\right) .
\eqno (18)$$

\nn
We have thus described the Hilbert space
structure of $\Ha$ simply in terms of the
canonical symplectic form it carries and
the fundamental decomposition (15).
[Here, and henceforth, we will let $f_{\pm}$
denote the projection of $f$ to $W_{\pm}$, etc.].

\nn
{\bf $\Ha$ and the Dirichlet space on the disc:}
In order to survey our work on the universal period map,
we have to rely on interpreting the functions in $H^{1/2}$
as boundary values (``traces'') of
functions in the disc $\D$ that have finite
Dirichlet energy, (i.e. the first
derivatives are in $L^2 (\D)$).

Define the following ``Dirichlet space'' of
harmonic functions:

$$\Da =
\{ F : \D \ra \RR : F \
{\mathop{\rm  is \ harmonic \ }\nolimits},
F(0) = 0, \ \and \ E(F) < \Inf \}
\eqno (19)$$

\nn
where the energy $E$ of any (complex-valued) $C^1$ map on $\D$
is defined as the $L^2 (\D)$ norm of $\grad (F)$ :

 $$
\left\Vert{ F }\right\Vert_{\Da}^{2} = E(F) =
{1\over 2 \pi }
\int_{ }^{ }\!
\int_{ \D}^{ }
\left({ \left\vert{ { \part F \over
\part x }  }\right\vert_{}^{2} +
\left\vert{ { \part F \over
 \part y }  }\right\vert_{}^{2}
}\right) dx dy
\eqno (20)$$

\nn
$\Da$, and its complexification $\Da_\CC$,
 will be Hilbert spaces with respect to
this energy norm.

We want to identify the space $\Da$ as precisely
the space of harmonic functions in
$\D$ solving the Dirichlet problem for functions
in $\Ha$. Indeed, the {\it
Poisson integral representation allows us
to map $P: \Ha \ra \Da$ so that $P$ is
an isometric isomorphism of Hilbert spaces.}

To see this let
$f (e^{i \t}) = \ds
\sum_{ - \Inf }^{ \Inf } u_n e^{i n \t}$
be an arbitrary member of
$\Ha_\CC$. Then the Dirichlet extension of
$f$ into the disc is:

$$ F(z) =
\sum_{n =  - \Inf }^{ \Inf }
u_n r^{|n|} e^{i n \t } =
\left({ \sum_{n = 1 }^{ \Inf } u_n z^{n} }\right) +
\left({ \sum_{m = 1 }^{ \Inf } u_{- m } \oo z^{m}  }\right)
\eqno (21)$$

\nn
where $z = re^{i \t}$.
 From the above series one can directly compute the $L^2 (\D)$
norms of $F$ and also of $\grad (F) = (\part F/ \part x, \part F/ \part y)$.
One obtains the following:

$$E(F) =
{1 \over 2 \pi }
\int_{ }^{ }\!
\int_{ \D}^{ }
| \grad (F) |^2 =
\sum_{ - \Inf }^{ \Inf } | n | |u_n |^2 \equiv
\left\Vert{ f }\right\Vert_{\Ha}^{2}
< \Inf
\eqno (22)$$

We will require crucially the well-known formula of Jesse Douglas
expressing the above energy of $F$ as the double integral on
$S^1$ of the square of the first differences of the boundary values $f$.

$$
E(F)=
{1\over {16{\pi}^2}}
\int_{S^1}^{}\!
\int_{S^1}^{}
[{(f(e^{i\t})-f(e^{i\p}))}/{sin({(\t - \p)}/2)}]^2 d\t d\p
\eqno(23)
$$

Transferring to the real line by the M\"obius transform identification
of $\Ha$ with $H^{1/2}(\RR)$ as explained before, the above identity
becomes as simple as:

$$
E(F)=
{1\over {4{\pi}^2}}
\int_{\RR}^{}\!
\int_{\RR}^{}
[{(f(x)-f(y))}/{(x-y)}]^2 dxdy=
\left\Vert{ f }\right\Vert^{2}
\eqno(24)
$$

\nn
Calculating from the series (21), the $L^2$-norm of $F$ itself is:

$$
{1 \over 2 \pi }
\int_{ }^{ }\!
\int_{\D }^{ }
| F|^2 dx dy
= \sum_{ - \Inf }^{ \Inf }
{  |u_n |^2  \over (|n| + 1)}
\leq E(F) < \Inf
\eqno (25)$$

\nn
(22) shows that indeed
{\it Dirichlet extension is isometric from }
$\Ha$ {\it to} $\Da$, whereas (25) shows
that the functions in $\Da$ are themselves in
$L^2$, so that the {\it the inclusion
of  $\Da \hra L^2 (\D)$ is continuous}.
(Bounding the $L^2$ norm of $F$ by the
$L^2$ norm of its derivatives is  a ''Poincar\'e
inequality'').

It is therefore clear that $\Da$
is a subspace of the usual Sobolev space $H^1
(\D)$ comprising those functions
in $L^2 (\D)$ whose first derivatness (in the
sense of distributions) are also in $L^2 ( \D)$.
The theory of function spaces implies
(by the ``trace theorems'') that $H^1$ functions
lose half a derivative in going to
a boundary hyperplane. Thus it is known that the
functions in $\Da$ will indeed have
boundary values in $H^{1/2}$.

Moreover, the identity (24) shows that
for any $F \in \Da$, the Fourier expansion of
the trace on the boundary circle is a Fourier series with
$\sum |n| |u_n|^2 < \Inf$.
But Fourier expansions with coefficients in such a
weigted $\ell_2$ space, as in our situation,
 are known to converge {\it
quasi-everywhere} (i.e. off a set of logarithmic
capacity zero) on $S^1$. See
Zygmund [Z, Vol 2, Chap. XIII]. The identification between $\Da$ and
$\Ha$ (or $\Da_\CC$ and $\Ha_\CC$) is now complete.

It will be necessary for us to identify the
 $W_\pm$ polarization of $\Ha_\CC$ at
the level $\Da_\CC$. In fact, let us decompose
the harmonic function $F$ of (21)
into its holomorphic and anti-holomorphic parts ;
these are $F_+ $ and $F_-$,
which are (respectively) the two sums bracketed
 separately on the right hand side
of (21). Clearly $F_+$ is a holomorphic
function extending $f_+$ (the $W_+$ part
of $f$), and $F_-$ is anti-holomorphic extending
 $f_-$. We are thus led to
introduce the following space of holomorphic functions
whose derivatives are in $L^2 (\D)$:

$$
\Hol_2 (\D) =
\{ H: \D \ra \CC: H \
{\mathop{\rm is \ holomorphic \  }\nolimits},
H(0) = 0
\ \
\and
\ \
\int_{ }^{ }\!
\int_{\D }^{ }
| H' (z) |^2 dx dy < \Inf \} .\eqno (26)$$

\nn
This is a complex Hilbert space with the norm
$$
\left\Vert{ H }\right\Vert_{}^{2}
 = {1 \over 2 \pi }
\int_{ }^{ }\!
\int_{ \D }^{ }
| H' (z)  |^2 dx dy .
\eqno (27)$$

\nn
If $H(z) = \ds \sum_{ n = 1}^{\Inf } u_n z^n$,
 a computation in polar coordinates
(as for (21), (25)) produces

$$
\left\Vert{ H }\right\Vert_{}^{2} =
\sum_{ n = 1}^{\Inf }
n \left\vert{ u_n }\right\vert_{}^{2}.
\eqno (28)$$

\nn
Equations (27) and (28) show that the norm-squared
is the Euclidean area of the
(possibly multi-sheeted) image of the map $H$.

We let $\oo {\Hol_2} (\D)$ denote the Hilbert space
of antiholomorphic functions
conjugate to those in $\Hol_2 (\D)$. The norm is defined by
stipulating that the anti-linear isomorphism of $\Hol_2$ on
$\oo {\Hol_2} $ given by conjugation should be an isometry. The
Cauchy-Riemann equation for $F_+$ and $\oo F_-$ imply that

$$
\left\vert{ \grad (F) }\right\vert_{}^{2}
= 2 \left\lbc{ \left\vert{ F'_+ }\right\vert_{}^{2}
- \left\vert{ F'_- }\right\vert_{}^{2} }\right\rbc
\eqno (29)$$

\nn
and hence
$$
\left\Vert{ F_+ }\right\Vert_{}^{2}
+ \left\Vert{ F_- }\right\Vert_{}^{2} =
\left\Vert{ f }\right\Vert_{\Ha_\CC}^{2}.
\eqno (30)$$

Now, the relation between $\Da$ (harmonic functions in
$H^1 (\D))$ and $\Hol_2 (\D)$ is transparent, so that the
holomorphic functions in $\Hol_2$ will have non-tangential
limits quasi-everywhere on $S^1$ - defining a function $W_+$.
We thus collect together, for the record,
the various representations of our basic Hilbert space:

\noindent
{\bf THEOREM 3.1: }
There are {\it canonical isometric isomorphisms} between the
following complex Hilbert spaces:

\medskip
(1) $\Ha_\CC = H^{1/2} \left({ S^1, \CC }\right) / \CC
= \CC \ot H^{1/2}(\RR)
= W_+ \op W_-$;

\medskip
(2) The sequence space $\ell_2^{1/2} (\CC)$ (constituting the Fourier
coefficients of the above quasi-everywhere defined functions);

\medskip
(3) $\Da_\CC$, comprising normalized finite-energy harmonic
functions (either on $\D$ or on the half-plane $U$); [the norm-squared
being given by (20) or (22) or (23) or (24)];

\medskip
(4) $\Hol_2 (\D) \op \oo {\Hol_2} (\D)$.

\medskip
\nn
Under the canonical identifications, $W_+$ maps to
$\Hol_2 (\D)$ and $W_-$ onto $ \oo {\Hol_2} (\D)$.
\xx

\noindent
{\bf Remark:}
Since the isomorphisms of the Theorem are all isometric, and
because the norm arose from the canonical symplectic structure,
(formulas (16), (17), (18)), it is possible and
instructive to work out the
formulas for the symplectic form $S$ on $\Da_\CC$ and on
$\Hol_2 (\D) $.

\medskip
\nn
{\bf 4.Quasisymmetric invariance:}

Quasiconformal (q.c.) self-homeomorphisms of the disc $\D$ (or
the upper half-plane) $U$ are known to extend continuously to
the boundary. The action on the boundary circle (respectively,
on the real line $\RR$) is called a {\it quasisymmetric}
(q.s.) homeomorphism. Now, $\vp: \RR \ra \RR$
is quasisymmetric if and only if, for all $x \in
\RR$ and all $t > 0$, there exists some $K > 0$ such that

$$
{1 \over K}
\leq { \vp (x + t) - \vp (x) \over  \vp (x) - \vp (x - t) }
\leq K \eqno (31)
$$
\noindent
On the circle this condition for $\vp: S^1 \ra S^1$ means
that $|\vp(2I)|/|\vp(I)| \leq K$, where $I$ is any interval on $S^1$
of length less that $\pi$, $2I$ denotes the interval obtained
by doubling $I$ keeping the same mid-point, and $| \bu |$
denotes Lebesgue measure on $S^1$.

Given any orientation preserving homeomorphism $\vp: S^1 \ra
S^1$, we use it to pullback functions in $\Ha$ by
precomposition:
$$V_\vp (f) = \vp^* (f) =
f \circ \vp -
{1 \over 2 \pi}
\int_{S^1 }^{ }
(f \circ \vp).
\eqno (32)$$
{\it This is the basic symplectic operator on $\Ha$ associated to
any quasisymmetric homeomorphism of $S^1$ that produces the universal
period mapping on $T(\Delta)$.}

\smallskip
\noindent
{\bf THEOREM 4.1:}
$V_\vp$ maps $\Ha$ to itself (i.e., the space $\Ha \circ
\vp$ is $\Ha$) if and only if $\vp$ is quasisymmetric.
The operator norm of $V_\vp \leq {\sqrt{K + K^{-1}}}$, whenever $\vp$
allows a K-quasiconformal extension into the disc.

\smallskip
\noindent
{\bf COROLLARY 4.2:}
The group of all quasisymmetric homeomorphism on $S^1, QS
\left({ S^1 }\right)$, acts faithfully by bounded toplinear
automorphisms on the Hilbert space $\Ha$ (and therefore also
on $\Ha_\CC$).

\medskip
\noindent
{\bf Proof of sufficiency:}
Assume $\vp$ is q.s. on $S^1$, and let $\Phi:\D \ra \D$ be
any quasiconformal extension. Let $f \in \Ha$ and suppose
$P(f) = F \in \Da$ is its unique harmonic extension into $\D$.
Clearly $ G \equiv F \circ \Phi$ has boundary values $f \circ
\vp$, the latter being (like $f$) also a continuous function
on $S^1$ defined quasi-everywhere. [Here we recall that q.s.
homeomorphisms carry capacity zero sets to again such sets,
although measure zero sets can become positive measure.]
To prove that $f \circ \vp$ minus its mean value is in $\Ha$,
it is enough to prove that the Poisson integral of $f \circ
\vp$ again has finite Dirichelet energy. Indeed we will show

$$
E({\mathop{\rm  harmonic \ extension \ of }\nolimits}
\ \vp^* (f)) \leq 2
\left({ {1 + k^2 \over 1 - k^2} }\right) E (F).
\eqno (33)$$

\noindent
Here $0 \leq k < 1$ is the q.c. constant for $\Phi$, i.e.,

$$\left\vert{ \Phi_{\oo z} }\right\vert_{}^{}
\leq k
\left\vert{  \Phi_{ z} }\right\vert_{}^{}
\ \ {\mathop{\rm
a.e. \ in  }\nolimits}  \ \D .
$$

\noindent
The operator norm of $V_\vp$ is thus no more that
$2^{1/2 }
\left({ {\ds 1 + k^2 \over \ds 1 - k^2} }\right)^{1/2 }$. The last
expression is equal to the bound quoted in the Theorem, where, as usual,
$ K =  (1+k)/(1-k)$.

Towards establishing (33) we prove that the inequality holds
with the left side being the energy of the map $G = F \circ
\Phi$. Since $G$ is therefore also a finite energy extension
of $f \circ \vp $ to $\D$, {\it Dirichlet's principle}
(namely that the minimal energy amongst all extensions
is achieved by the harmonic (Poisson integral) extension)
implies the required inequality.\xx

\noindent
{\bf Remark:}
Since the Dirichlet integral in two dimensions is
invariant under conformal mappings, it is
not too surprising that it is quasi-invariant
under quasiconformal transformations. Such quasi-invariance
has been noted and applied before. See [A1].

\noindent
{\bf Proof of necessity:}
The idea of this proof is taken from the notes of M. Zinsmeister.
We express our gratitude for his generosity.

Since two-dimensional Dirichlet integrals are conformally
invariant, we will pass to the upper half-plane $U$ and its
boundary line $\RR$ to aid our presentation. As explained earlier, using
the Cayley transform we transfer
everything over to the half-plane; the traces
on the boundary constitute the space of quasi-everywhere defined
functions called  $H^{1/2} (\RR)$.

 From the Douglas identity, equation (24), we recall
that an {\it equivalent way of expressing the Hilbert space norm on }
$H^{1/2} (\RR)$ is

$$
\left\Vert{ g }\right\Vert_{}^{2} =
{1 \over {4 {\pi^2}}} \int_{ }^{ }
\!\int_{\RR^2 }^{ } \left\lbk{
{ g(x) - g(y) \over x - y }  }\right\rbk_{}^{2}dxdy,
{}~~g \in H^{1/2} (\RR).
\eqno (34)$$

\noindent
Equation (34) immediately shows that
 $ \left\Vert{ g }\right\Vert = \left\Vert{ \ww g }\right\Vert$
where $\ww g(x) = g(ax + b)$ for any real $a (\ne 0) $ and
$b$. This will be important.

Assume that $\vp: \RR \ra \RR$ is an orientation preserving
homeomorphism such that
$V_{\vp^{-1}}: H^{1/2} (\RR) \ra H^{1/2} (\RR)$ is a bounded
automorphism.
Let us say that the norm of this operator is $M$.

Fix a bump function $f \in C_0^\Inf (\RR)$ such that $f \equiv
1$ on $[- 1, 1] f \equiv 0$ outside $[- 2, 2]$ and $0 \leq f
\leq 1$ everywhere. Choose any $c \in \RR$ and any positive
$t$. Denote $I_1 = [x - t, x]$ and $I_2 = [x, x + t]$. Set
$g(u) = f(au + b)$, choosing $a$ and $b$ so that $g$ is
identically $1$ on $I_1$ and zero on $[x + t, \Inf)$.

By assumption, $g \circ \vp^{- 1}$ is in $H^{1/2} (\RR)$ and
$
\left\Vert{ g \circ \vp^{-1} }\right\Vert_{}^{}
\leq M \left\Vert{ g }\right\Vert_{}^{} =
M \left\Vert{ f }\right\Vert_{}^{}$.
We have

$$
\eqalignno{
 M \left\Vert{ f }\right\Vert_{}^{}
&\geq
\int_{ }^{ }
\!\int_{\RR^2 }^{ }
\left\lbk{
{ g \circ \vp^{- 1}(u) -
g \circ \vp^{- 1}(v) \over u - v }  }\right\rbk_{}^{2}
du dv\cr
&\geq
\int_{ v = \vp (x - t) }^{ v = \vp (x)}
\int_{ u = \vp (x + t) }^{ u = \Inf}
{1 \over (u - v)^2 } du dv\cr
&=
\log
\left({ 1 +  {  \vp (x) - \vp (x - t) \over
\vp (x + t) - \vp (x)  } }\right)  .
&(35) \cr}$$

\nn
[We have utilised the elementary integration
$\ds \int_{\g}^{ \Inf} \ds \int_{ \a }^{ \b}
{1 \over (u - v)^2 } du dv =
\log \left({ 1 + {\b - \a \over \g - \b } }\right) $, for
$\a<\b<\g$.]
We thus obtain the result that

$${  \vp (x + t ) - \vp (x ) \over
\vp (x ) - \vp (x - t)  }
\geq
{ 1 \over e^{M \left\Vert{ f }\right\Vert } - 1 }
$$

\nn
for arbitrary real $x$ and positive $t$. By utilising
symmetry, namely by shifting the bump to be $1$ over $I_2$ and
$0$ for $u \leq x - t$, we get the opposite inequality:

$$
{  \vp (x + t ) - \vp (x ) \over
\vp (x ) - \vp (x - t)  }
\leq  e^{M \left\Vert{ f }\right\Vert } - 1.
$$

\nn
The Beurling-Ahlfors condition on $\vp$ is verified, and we
are through. Both the theorem and its corollary are proved. \xx



\bigskip
\nn
{\bf 5.The invariant symplectic structure:}

The quasisymmetric homeomorphism group,
$QS \left({  S^1}\right)$, acts on $\Ha$ by precomposition
(equation (32)) as bounded operators, {\it preserving the
canonical symplectic form} $S: \Ha \ts \Ha \ra \RR$.
This is the central fact which we will now expose; it
is the crux on which the definition of the period
mapping on all of $T(1)$ hinges:

\nn
{\bf PROPOSITION 5.1:}
For every $\vp \in QS \left({  S^1}\right)$, and all $f, g \in \Ha$,

$$
S \left({ \vp^* (f), \vp^* (g)  }\right) =
S (f, g). \eqno (36)$$

\nn
Considering the complex linear extension of the action to
$\Ha_\CC$, one can assert that the only quasisymmetrics which
preserve the subspace $W_+ = \Hol_2 (\D)$ are the M\"obius
transformations. Thus M\"ob $\left({  S^1}\right)$ acts as
{\it unitary} operators on $W_+$ (and $W_-$).

Before proving the proposition we would like to point out that
this canonical symplectic form enjoys a far stronger
invariance property. The proof is an exercise in calculus.

\noindent
{\bf LEMMA 5.2:}
If $\vp: S^1 \ra S^1$ is any (say $C^1$) map of winding
number (= degree) $k$, then

$$S(f \circ \vp, g \circ \vp) = kS (f, g) \eqno (37)$$

\nn
for arbitrary choice of ($C^1$) functions $f$ and $g$ on
the circle. In particular, $S$ is invariant under pullback by
all degree one mappings.

\noindent
{\bf Proof of Proposition:}
The Lemma shows that (36) is true whenever the quasisymmetric
homeomorphism $\vp$ is at least $C^1$. By
standard facts of quasiconformal mapping theory
we know that for arbitrary q.s $\vp$, there exist real analytic q.s.
homeomorphisms $\vp_m$ (with the same quasisymmetry constant
as $\vp$) that converge uniformly to $\vp$. An approximation
argument (see[NS]) then proves the required result.

If the action of $\vp$ on $\Ha_\CC$ preserves $W_+$ it is easy
to see that $\vp$ must be the boundary values of some
holomorphic map $\Phi: \D \ra \D$. Since $\vp$ is a
homeomorphism one can see that $\Phi$ is  a holomorphic
homeomorphism (as explained in [N1])
-- hence a M\"obius transformation. Since every $\vp$ preserves
$S$, and since $S$ induces the inner product on $W_+$ and
$W_-$ by (16) (17), we note that such a symplectic
transformation preserving $W_+$ must necessarily act unitarily.
\xx

\medskip
\nn
{\bf Motivational remark:}
Theorem 4.1 and Proposition 5.1 enable us to consider $QS
\left({  S^1}\right)$ as a subgroup of the bounded symplectic
operators on $\Ha$. Since the heart of the matter in extending
the period mapping from Witten's homogeneous space $M$ (as in
[N1], [N2]) to $T(1)$ lies in the property of preserving this
symplectic form on $\Ha$, we must prove that $S$ is indeed the
{\bf unique} symplectic form that is $\Diff
\left({  S^1}\right)$ or $QS \left({  S^1}\right)$ invariant.
It is all the more surprising that the form $S$ is canonically
specified by requiring its invariance under simply the
3-parameter subgroup M\"ob
$\left({  S^1}\right) (\hra \Diff \left({  S^1}\right)
\hra QS\left({  S^1}\right)$ ). We sketch a proof of this
uniqueness result:

\noindent
{\bf THEOREM 5.3:}
Let $S \equiv S_1$ be the canonical symplectic form on $\Ha$.
Suppose $S_2: \Ha \ts \Ha \ra \RR$ is any other continuous
bilinear form such that
$S_2 ( \vp^* (f), \vp^* (g)) = S_2 (f, g)$, for all $f, g$ in
$\Ha$ whenever $\vp$ is in M\"ob
$\left({  S^1}\right) $. Then $S_2$ is necessarily a real
multiple of $S$. Thus every form on $\Ha$ that is M\"ob
$\left({  S^1}\right) \equiv PSL (2, \RR)$ invariant is
necessarily non-degenerate (if not identically zero) and
remains invariant under the action of the whole of $QS
\left({  S^1}\right)$. (Also, it automatically satisfies the
even stronger invariance property (37)).

The proof requires some representation theory.
We start with an easy lemma:

\noindent
{\bf LEMMA 5.4:}
The duality induced by canonical form $S_1$ is (the negative
of) the Hilbert transform (equation (6)). Thus the map $\Si_1$
(induced by $S_1$) from $\Ha$ to $\Ha^*$ is an invertible
isomorphism.\xx

{\it The basic tool in proving the uniqueness of the symplectic form
is to consider the ''intertwining operator'':}
$$M = \Si_1^{-1} \circ \Si_2: \Ha \ra \Ha \eqno (38)$$
\nn
which is a bounded linear operator on $\Ha$ by the above Lemma.

\noindent
{\bf LEMMA 5.5:}
$M$ commutes with every invertible linear operator on $\Ha$
that preserves both the forms $S_1$ and $S_2$.

\noindent
{\bf Proof:}
$M$ is defined by the identity $S_1 (v, M w) = S_2 (v, w)$. If
$T$ preserves boths forms then one has the string of equalities:

$$S_1 (Tv, TM w) =
S_1 (v, M w) =
S_2 (v,  w) =
S_2 (Tv, T w) =
S_1 (Tv, MT w) $$

\nn
Since $T$ is assumed invertible, this is the same as saying

$$ S_1 (v, TM w) =
S_1 (v, MT w) , \ \for \ \all \ v , w \in \Ha
\eqno (39)
$$

\nn
But $S_1$ is non-degenerate, namely $\Si_1$ was an
isomorphism. Therefore (39) implies that $TM \equiv MT$, as desired.
\xx

It is clear that to prove that $S_2$ is simply a real multiple of $S_1$
means that the intertwining  operator $M$ has to be just
multiplication by a scalar. This can now be deduced by looking
at the complexified representation of
M\"ob $\left({  S^1}\right)$ on $\Ha_\CC$, which is unitary,
and applying Schur's Lemma. [I am indebted to Graeme Segal and Ofer Gabber
for helpful conversations in this regard.]

\noindent
{\bf LEMMA 5.6:}
The unitary representation of $SL(2, \RR)$ on $\Ha_\CC$
decomposes into precisely two irreducible pieces - namely on
$W_+ $ and $W_-$. In fact these two representations correspond
to the two lowest (conjugate) members in the discrete series
for
$SL (2, \RR)$.

\noindent
{\bf Proof:}
We claim that the representation given by the operators
$V_\vp$ on $W_+$ (equation (32)), $\vp \in $ M\"ob
$\left({  S^1}\right)$, can be indentified with the ''$m = 2$'' case
of the discrete series of {\it irreducible} unitary
representations of $SL(2, \RR)$.
Note, M\"ob$\left({  S^1}\right) \equiv PSU(1, 1)
\cong SL(2, \RR)/(\pm I)$. Recall from
Theorem 3.1 that $W_+$ is identifiable as
$\Hol_2 (\D)$. The action of $\vp$ is given on $\Hol_2$ by :

$$V_\vp (F) = F \circ \vp - F \circ \vp (0), ~~F \in \Hol_2 (\D).
\eqno (39)$$

\nn
But $\Hol_2$ consists of normalized $(F(0) = 0) $ holomorphic
functions in $\D$ whose {\it derivative is in}
$L^2 (\D$, Euclidean measure). From (39), by the chain rule,

$$
{d \over dz } V_\vp (F) =
\left({ { dF \over dz } \circ \vp }\right)  \vp'
\eqno (40)$$

So we can rewrite the representation on the derivatives of the
functions in $\Hol_2$ by the formula (40) - which coincides
with the standard formula for the discrete series irreducible
representation at lowest index.

It is clear that the representation on the conjugate space will
correspond to the $m = - 2$ (highest weight vector of weight $
- 2$) case of the discrete series. In particular, the
representations we obtain of M\"ob $\left({  S^1}\right)$ by unitary
operators of $W_+ $ and $W_-$ are both {\it irreducible. }
The Lemma is proved. \xx

\noindent
{\bf Proof of Theorem 5.3:}
By Lemma 5.5, (the $\CC$-linear extension of ) the
intertwining operator $M$ commutes with every one of the
unitary operators $V_\vp : \Ha_\CC \ra \Ha_\CC$ as $\vp$
varies over M\"ob $\left({  S^1}\right)$.
Since $W_+$ and $W_-$ are the only two invariant subspaces for
all the $V_\vp$, as proved above, it follows that $M$ must map
$W_+$ either to $W_+$ or to $W_-$. Let us first assume the
former case. Then $M$ commutes with all the unitary operators
$V_\vp$ on $W_+$, which we know to be an irreducible
representation. {\it Schur's Lemma} says that a unitary
representation will be irreducible if and only if the only
operators that commute with all the operators in
the representation are simply the scalars.
Since $M$ was a real operator to start with, that
scalar must be real.
The alternative assumption that $M$ maps $W_+$ to $W_-$ is
easily verified to be untenable. \xx

\nn
{\bf Remark:}
Our proof of absolute naturality of the symplectic
form is complete. We will utilise Theorem 5.3 in understanding the
$H^{1/2}$ space as a {\it Hilbertian} space, -- namely a space
possessing a fixed
symplectic structure but a large family of compatible complex
structures. That is the nature of a cohomology
space on a Riemann surface as the complex structure
on the surface varies -- but the topology -- which
determines the cup-product symplectic form, remains invariant.

\bigskip
\nn
{\bf 6.The $H^{1/2}$ space as first cohomology:}

The Hilbert space $H^{1/2}$, that is the hero of our tale, can be
interpreted as the first cohomology space with real coefficients of the
"universal Riemann surface" -- namely the unit disc -- in a
Hodge-theoretic sense. That will be fundamental for us in explaining the
properties of the period mapping on the universal Teichm\"uller space.

In fact, in the classical theory of the period mapping,
the vector space $H^1(X,\RR)$
plays a basic role, $X$ being a closed orientable topological
surface of genus $g$ to start with. This real vector space comes
equipped with a canonical symplectic structure given by the cup-product
pairing, $S$. Now, whenever $X$ has a complex manifold structure, this
real space $H^1(X,\RR)$ of dimension $2g$ gets endowed with {\it a
complex structure $J$ that is compatible with the cup-pairing $S$}. This
happens as follows: When $X$ is a Riemann surface, the cohomology space
above is precisely the vector space of real harmonic 1-forms on $X$, by
the Hodge theorem. Then the {\it complex structure $J$ is the Hodge
star operator on the harmonic 1-forms}. The compatibility with the cup
form is encoded in the relationships (41) and (42):

$$
S(J\a, J\b)= S(\a, \b),
{}~~~{\rm for~~ all}~~ \a , \b \in  H^1(X,\RR)
\eqno(41)
$$
\noindent
and that, intertwining $S$ and $J$ exactly as in equation (11),
$$
S(\a, J\b) = inner ~~ product(\a, \b)
\eqno(42)
$$
should define a positive definite inner product on $H^1(X,\RR)$.
[In fact, as we will further describe in Section 8, the Siegel disc of
period matrices for genus $g$ is precisely the space of all the
$S$-compatible complex structures $J$.] {\it Consequently,
the period mapping
can be thought of as the variation of the Hodge-star complex structure
on the topologically determined symplectic vector space $H^1(X,\RR)$.}

\nn
{\bf Remark:} Whenever $X$ has a complex structure, one gets an isomorphism
between the real vector space  $H^1(X,\RR)$ and
the $g$ dimensional complex vector space
$H^1(X,\cal O)$, where $\cal O$ denotes the sheaf of germs of holomorphic
functions. That is so because $\RR$ can be considered as a subsheaf of
$\cal O$ and hence there is an induced map on cohomology.
It is interesting to check that this natural map is an
isomorphism, and that the complex structure so induced on
$H^1(X,\RR)$ is the same as that given above by the Hodge star.

For our purposes it therefore becomes relevant to consider, for an {\it
arbitrary} Riemann Surface $X$, {\it the Hodge-theoretic first
cohomology vector space as the space of $L^2$ (square-integrable) real
harmonic 1-forms on $X$}. This real Hilbert space will be denoted
$\Ha(X)$. Once again, in complete generality, this Hilbert space has a
non-degenerate symplectic form $S$ given by the cup (= wedge) product:

$$
S(\p_1, \p_2) =
\int _{}^{}\!
\int_{X}^{}
\p_1 \wedge \p_2
\eqno(43)
$$

\nn
and the Hodge star is the complex structure $J$ of $\Ha(X)$ which is
again compatible with $S$ as per (41) and (42). In fact,
one verifies that the $L^2$ inner
product on $\Ha(X)$ is given by the triality
relationship (42) -- which is the same as (11).

Since in the universal Teichm\"uller theory we deal with the "universal
Riemann surface" -- namely the unit disc $\D$ -- (being the universal
cover of all Riemann surfaces), we require the

\noindent
{\bf PROPOSITION 6.1:} For the disc $\D$, the Hilbert space $\Ha(\D)$ is
isometrically isomorphic to the real Hilbert space $\Ha$ of Section 3.
Under the canonical identification the cup-wedge pairing is the
canonical symplectic form $S$ and the Hodge star becomes the
Hilbert-transform on $\Ha$.

\noindent
{\bf Proof:} For every  $\p \in \Ha(\D)$
there exists a unique real harmonic
function $F$ on the disc with $F(0)= 0$  and $dF = \p$. Clearly  then,
$\Ha(\D)$ is isometrically isomorphic to the Dirichlet space $\Da$ of
normalized real harmonic functions having finite energy. But in Section
3 we saw that this space is isometrically isomorphic to $\Ha$ by passing
to the boundary values of $F$ on $S^1$.

If $\p_1 = dF_1$ and $\p_2 = dF_2$ , then integrating $\p_1 \wedge \p_2$
on the disc amounts to, by Stokes' theorem,
$$
\int_{}^{}\!\int_{\D}^{}  dF_1 \wedge dF_2 =
\int_{S^1}^{} F_1 dF_2 = S(F_1, F_2)
$$
as desired.

Finally, let $\p = udx + vdy$ be a $L^2$ harmonic 1-form with $\p = dF$.
Suppose $G$ is the harmonic conjugate of $F$ with $G(0)=0$.
Then $dF + idG$ is a holomorphic 1-form on $\D$ with real part $\p$. It
follows that the Hodge star maps $\p$ to $dG$; hence, under the above
canonical identification of $\Ha(\D)$ with $\Ha$, we see that the
star operator becomes the Hilbert transform. \xx

\bigskip
\nn
{\bf 7.Quantum calculus and $H^{1/2}$:}

\nn
A.Connes has proposed (see, for example,
[CS] and Connes' book "Geometrie Non-Commutatif")
a "quantum calculus" that associates to a
function $f$ an operator that should
be considered its quantum derivative -- so that the operator theoretic
properties of this $d^{Q}(f)$ capture the smoothness properties of the
function.  One advantage is that this operator can undergo all the
operations of the functional calculus. The fundamental definition in one
real dimension is

$$
d^{Q}(f)=[J,M_{f}]
\eqno(44)
$$

\nn
where $J$ is the Hilbert transform in one dimension explained in
Section 3, and $M_{f}$ stands for (the generally unbounded) operator
given by multiplication by $f$. One can think of the quantum derivative
as operating (possibly unboundedly) on the Hilbert space $L^{2}(S^1)$ or
on other appropriate function spaces.

\medskip
\nn
{\bf Note:} We
will also allow quantum derivatives to be taken with respect to other
Hilbert-transform like operators; in particular, the Hilbert transform
can be replaced by some conjugate of itself by a suitable automorphism
of the Hilbert space under concern. In that case we will make explicit
the $J$ by writing $d^{Q}_{J}(f)$ for the quantum derivative. See
below for applications.

As sample results relating the properties of the quantum derivative
with the nature of $f$, we quote:
$d^{Q}(f)$ is a bounded operator on $L^{2}(S^1)$ {\it if and only if}
the function $f$ is of bounded mean oscillation. In fact, the operator
norm of the quantum derivative is equivalent to the BMO norm of $f$.
Again, $d^{Q}(f)$ is a compact operator on $L^{2}(S^1)$ {\it if and only if}
$f$ is in $L^{\Inf}(S^1)$ and has vanishing mean oscillation.
Also, if $f$ is H\"older, (namely in some H\"older class), then
the quantum derivative acts as a compact operator on H\"older.
See [CS], [CRW].
Similarly, the requirement that $f$ is a member of certain Besov spaces
can be encoded in properties of the quantum derivative.

Our Hilbert space $H^{1/2}(\RR)$ has a very simple
interpretation in these terms:

\nn
{\bf PROPOSITION 7.1:} $f \in H^{1/2}(\RR)$  if and only if the operator
$d^{Q}(f)$ is Hilbert-Schmidt on $L^{2}(\RR)$ [or on $H^{1/2}(\RR)$].
The Hilbert-Schmidt norm
of the quantum derivative {\it coincides} with the $H^{1/2}$ norm of
$f$.

\nn
{\bf Proof:} Recall that the Hilbert transform on the real line is given
as a singular integral operator with integration kernel $(x-y)^{-1}$.
A formal calculation therefore shows that

$$
(d^{Q}(f))(g)(x) =
\int_{\RR}^{} {{f(x)-f(y)}\over {x-y}} g(y)dy
\eqno(45)
$$

But the above is an integral operator with kernel $K(x,y)=
(f(x)-f(y))/(x-y)$, and such an operator is Hilbert-Schmidt if and only
if the kernel is square-integrable over $\RR^{2}$. Utilising now the
Douglas identity -- equation (24) -- we are through. \xx

Since the Hilbert transform, $J$, is the standard complex structure on
the $H^{1/2}$ Hilbert space, and since this last space was shown to
allow an action by the quasisymmetric group, $QS(\RR)$, some further
considerations become relevant. Introduce the operator $L$ on 1-forms on
the line to functions on the line by:

$$
(L\vp)(x) =
\int_{\RR}^{} [log \vert x-y \vert]\vp (y) dy
\eqno(46)
$$

One may think
of the Hilbert transform $J$ as operating on
either the space of functions or on the space of 1-forms (by
integrating against the kernel $dx/(x-y)$). Let $d$ as usual denote
total derivative (from functions to 1-forms). Then notice that $L$ above
is an operator that is essentially a smoothing inverse of the
exterior derivative. In fact, $L$ and $d$ are connected to $J$
via the relationships:

$$
d \circ L = J_{1-forms}; ~~~ L \circ d = J_{functions}
\eqno(47)
$$

\medskip
\nn
{\bf The Quasisymmetrically deformed operators:} Given any
q.s. homeomorphism $h \in QS(\RR)$ we think of it as producing
a q.s. change of structure on the line, and hence we define the
corresponding transformed operators, $L^{h}$ and $J^{h}$ by
$L^{h}= h \circ L \circ h^{-1}$ and $J^{h}= h \circ J \circ h^{-1}$.
($J$ is being considered on functions in $\Ha = H^{1/2}(\RR)$,
as usual.)
The q.s homeomorphism (assumed to be say $C^1$ for the deformation
on $L$), operates standardly on functions and forms by pullback.
Therefore, {\it $J^{h}$ simply stands for the Hilbert transform
conjugated by the symplectomorphism $T_{h}$ of $\Ha$ achieved by
pre-composing by the q.s. homeomorphism $h$.} $J^{h}$ is thus a
new complex structure on $\Ha$.

\smallskip
\nn
{\bf Note:} The complex structures on $\Ha$ of type
$J^{h}$ are exactly those that constitute the image of $T(1)$ by the
universal period mapping. (See Section 8.)
The target manifold, the universal Siegel space,
can be thought of as a space of $S$-compatible complex
structures on $\Ha$.

Let us write the perturbation achieved by $h$ on these operators as the
"quantum brackets":

$$
\{h,L\}=L^{h} - L ;~~~  \{h,J\}=J^{h} - J.
\eqno(48)
$$

Now, for instance, the operator $d \circ \{h,J\}$ is represented
by the kernel $(h \times h)^{*}m - m$ where
$m = dx dy/{(x-y)^{2}}$. For
$h$ suitably smooth this is simply
$d_{y}d_{x}(log [({h(x)-h(y)})/({x-y})])$.
It is well known that
$(h \times h)^{*}m = m$ when $h$ is a M\"obius transformation.
Interestingly, therefore,
on the diagonal ($x=y$), this kernel
becomes ($1/6$ times) the Schwarzian derivative of $h$ (as a
quadratic differential on the line). For the other operators in
the table below the kernel computations are even easier.

Set $K(x,y) = log [({h(x)-h(y)})/({x-y})]$ for convenience.
We have the following table of quantum calculus formulas:

$$
\matrix {{\bf Operator} & {\bf Kernel} & {\bf On~~ diagonal} &
{\bf Cocycle~~on}~~QS(\RR) \cr
\{ h,L\} &  K(x,y) & log (h^\prime) & function-valued \cr
d \circ \{h,L\} & d_{x}K(x,y) & {h'' \over {h'}}dx & 1-form-valued \cr
d \circ \{h,J\} & d_{y}d_{x}K(x,y) & {1 \over 6}Schwarzian(h)dx^{2} &
quadratic-form-valued}
\eqno(49)
$$

\nn
{\it The point here is that these operators make sense when $h$ is merely
quasisymmetric}. If $h$ happens to be appropriately smooth, we can
restrict the kernels to the diagonal to obtain the respective
nonlinear classical derivatives (affine Schwarzian, Schwarzian, etc.)
as listed in the table above.

\bigskip
\nn
{\bf 8.The universal period mapping on $T(1)$:}

Having now all the necessary background results behind us, we are
finally  set to move into the theory of the universal period (or
polarisations) map itself.

As we said at the very start, the
Frechet Lie group, $Diff(S^1)$ operating by pullback (=
pre-composition) on smooth functions, has a faithful representation by
bounded symplectic operators on the symplectic vector space $V$ of
equation (1). That induced the natural map $\Pi$ of the homogeneous space
$M=Diff(S^1)/M\ddot{o}b(S^1)$ into Segal's version of the Siegel space of
period matrices.  In [N1] [N2] we had shown that this map:

$$
\Pi: Diff(S^1)/M\ddot{o}b(S^1) \ra Sp_{0}(V)/U
\eqno(50)
$$
\nn
is {\it equivariant, holomorphic, K\"ahler isometric immersion}, and
moreover that it qualifies as a
{\it generalised period matrix map}. Remember that the domain is a
complex submanifold of the universal space of Riemann surfaces $T(1)$.
Here $Sp_{0}$ denotes (see [S]) the symplectic automorphisms of Hilbert-
Schmidt type. The canonical Siegel symplectic-invariant metric exists
on the target space in (50).

 From the results of Sections 2, 3, and 4, 5, we know that the full
quasisymmetric group, $QS(S^1)$ operates as bounded symplectic operators
on the Hilbert space $\Ha$ that is the completion of the pre-Hilbert
space $V$. We also demonstrated
that the subgroup of $QS$ acting unitarily is the M\"obius subgroup.
Clearly then we have obtained the {\it extension of } $\Pi$
(also called $\Pi$ to save on nomenclature) {\it to the entire
universal Teichm\"uller space}:

$$
\Pi: T(1) \ra Sp(\Ha)/U
\eqno({\oo 50})
$$
\nn
{\bf Universal Siegel (period matrix) space:}
Let us first exhibit the nature of the complex Banach manifold that is
the target space of the period map (71). This space, which is the
universal Siegel period matrix space, denoted
$\cal S_{\infty}$, has several interesting descriptions:

\medskip
\nn
{\bf (a):}
$\cal S_{\infty}$=
{the space of positive polarizations of the symplectic Hilbert space
$\Ha$ }. Recall that a
positive polarization signifies the choice of
a closed complex subspace $W$ in $\Ha_{\CC}$ such that
(i) $\Ha_{\CC} = W \op \oo W$; (ii) $W$ is $S$-isotropic,
namely $S$ vanishes on
arbitrary pairs from $W$; and (iii) $iS(w, \oo w)$ defines the square of
a norm on $w \in W$. [In the classical genus $g$ situation (ii)
and (iii) are the bilinear relations of Riemann.]

\medskip
\nn
{\bf (b):}
$\cal S_{\infty}$=
{the space of $S$-compatible complex structure operators on $\Ha$ }.
That consists of bounded invertible operators $J$ of $\Ha$ onto itself
whose square is the negative identity and $J$ is compatible with $S$ in
the sense that requirements (41) and (42) are valid.
Alternatively, these are the complex structure operators $J$ on $\Ha$
such that $H(f,g) = S(f,Jg) + iS(f,g)$ is a positive definite Hermitian
form having $S$ as its imaginary part.

\medskip
\nn
{\bf (c):}
$\cal S_{\infty}$=
{the space of bounded operators $Z$ from $W_{+}$ to $W_{-}$ that satisfy
the condition of $S$-symmetry: $S(Z\a,\b)=S(Z\b,\a)$ and are in the unit
disc in the sense  that $(I-Z \oo Z)$ is positive definite}.
The matrix for $Z$ is the "period matrix" of the classical theory.

\medskip
\nn
{\bf (d):}
$\cal S_{\infty}$=
the homogeneous space $Sp(\Ha)/U$; here $Sp(\Ha)$ denotes all bounded
symplectic automorphisms of $\Ha$, and $U$ is the unitary subgroup
defined as those symplectomorphisms that keep the subspace $W_{+}$
(setwise) invariant.

\medskip
Introduce the {\it Grassmannian} $Gr(W_{+}, \Ha_{\CC})$ of subspaces of type
$W_{+}$ in $\Ha_{\CC}$, which is obviously a complex Banach manifold
modelled on the Banach space of all bounded operators from $W_{+}$ to
$W_{-}$. Clearly,  $\cal S_{\infty}$  is embedded in $Gr$ as a complex
submanifold.  The connections between the above descriptions of the
Siegel universal space are transparent:

\medskip
\nn
(a:b) the positive polarizing subspace $W$ is the $-i$-eigenspace of the
complex structure operator $J$ (extended to $\Ha_{\CC}$ by complex
linearity).

\medskip
\nn
(a:c) the positive polarizing subspace $W$ is the graph of the operator
$Z$.

\medskip
\nn
(a:d) $Sp(\Ha)$ acts transitively on the set of positive polarizing
subspaces. $W_{+}$ is a polarizing subspace, and the isotropy
(stabilizer) subgroup thereat is exactly $U$.

\medskip
\nn
{\bf $\Ha$ as a Hilbertian space:} Note that the method (b) above
describes the universal Siegel space as a space of Hilbert space
structures on the fixed underlying symplectic vector space $\Ha$. By the
result of Section 4 we know that the symplectic structure on $\Ha$ is
completely canonical, whereas each choice of $J$
above gives a Hilbert space inner
product on the space by intertwining $S$ and $J$ by the fundamental
relationship (11) (or (42)). Thus $\Ha$ is a {\it "Hilbertian space"},
which signifies a complete topological vector space with
a canonical symplectic structure but lots of compatible inner products
turning it into a Hilbert space in many ways.

\smallskip
\nn
We come to one of our {\it Main Theorems:}

\nn
{\bf THEOREM 8.1:} The universal period mapping $\Pi$ is an injective,
equivariant, holomorphic immersion between complex Banach manifolds.
Restricted to $M$ it is also an isometry between the canonical metrics.

\nn
{\bf Proof:}  From our earlier papers we know these facts
for $\Pi$ restricted to $M$. The proof of equivariance is the same
(and simple). The map is an injection because if we know the subspace
$W_{+}$ pulled back by $w_{\mu}$, then we can recover the q.s.
homeomorphism $w_{\mu}$. In fact, inside the given subspace look at
those functions which map $S^1$ homeomorphically on itself.
One sees easily that these must
be precisely the M\"obius transformations of the circle pre-composed
by $w_{\mu}$. The injectivity (global Torelli theorem) follows.

Let us write down the matrix for the symplectomorphism $T$ on $\Ha_{\CC}$
obtained by pre-composition by $w_{\mu}$. We will write in the standard
orthonormal basis $e^{ik\theta}/k^{1/2}$, $k=1,2,3..$ for $W_{+}$, and
the complex conjugates as o.n. basis for $W_{-}$.

In $\Ha_\CC = W_+ \op W_- $ block form, the matrix for operator
$T$ is given by maps $A: W_{+} \ra W_{+}$, and
$B:W_{-} \ra W_{+}$. The conjugates of $A$ and $B$ map $W_{-}$ to $W_{-}$,
and $W_{+}$ to $W_{-}$, respectively. The matrix entries for
$A=((a_{pq}))$ and $B=((b_{rs}))$ turn out to be:

$$
a_{pq}={(2\pi)^{-1}}{p^{1/2}}{q^{-1/2}}\int_{0}^{2\pi}
{(w_{\mu}(e^{i\theta}))^{q}}{e^{-ip\theta}}d\theta, ~~p,q \geq 1
$$
$$
b_{rs}={(2\pi)^{-1}}{r^{1/2}}{s^{-1/2}}\int_{0}^{2\pi}
{(w_{\mu}(e^{i\theta}))^{-s}}{e^{-ir\theta}}d\theta, ~~r,s \geq 1
$$

Recalling the standard action of symplectomorphisms on the Siegel disc
(model {\bf (c)} above), we see that the corresponding operator
[={\it period matrix}] $Z$ appearing from the Teichm\"uller point $[\mu]$
is given by:

$$
\Pi[\mu] = {\oo B}{A^{-1}}
$$
The usual proof of finite dimensions shows that for any symplectomorphism
$A$ must be invertible -- hence the above explicit formula makes sense.

Since the Fourier coefficients appearing in $A$ and $B$ vary only
{\it real-analytically} with $\mu$, it may be somewhat surprising that
$\Pi$ is actually {\it holomorphic}. In fact,
a computation of the first variation of $\Pi$ at the origin of $T(1)$
( i.e., the derivative map) in the Beltrami direction
$\nu$ shows that the following {\it Rauch variational formula} subsists:

$$
{(d\Pi([\nu]))_{rs}}	={{\pi}^{-1}}{(rs)^{1/2}}\int_{}^{}\int_{\Delta}^{}
\nu(z){z^{r+s-2}}dxdy
\eqno(51)
$$

\nn
The proof of this formula is as shown for $\Pi$ on the smooth
points submanifold $M$ in our earlier papers. The
manifest complex linearity of the derivative, i.e.,
the validity of the Cauchy-Riemann equations, combined with equivariance,
demonstrates that $\Pi$ is complex analytic on $T(1)$, as desired.

Utilizing the derivative (51) one shows that the Siegel symplectic
K\"ahler form pulls back on $M$ to the Weil-Petersson K\"ahler form.
\xx

\medskip
\nn
{\bf Interpretation of $\Pi$ as period map:}
Why does the map $\Pi$ qualify as a
universal version of the classical genus $g$ period maps? As we had
explained in our previous papers, in the light
of P.Griffiths' ideas, the classical period map may be thought of as
associating to a Teichm\"uller point a positive polarizing subspace of
the first cohomology $H^1(X,\RR)$. The point is that when $X$ has a
complex structure, then the complexified first cohomology decomposes as:
$$
H^1(X,\CC)=
H^{1,0}(X) \op H^{0,1}(X)
\eqno(52)
$$
\nn
The period map associates the subspace $H^{1,0}(X)$ -- which is positive
polarizing with respect to the cup-product symplectic form --
to the given complex structure on $X$.  Of course, $H^{1,0}(X)$
represents the holomorphic 1-forms on
$X$, and that is why this is nothing but the usual period mapping.

{\it But that is precisely what $\Pi$ is doing in the universal
Teichm\"uller space.} Indeed, by the results of Section 5, $\Ha$ is the
Hodge-theoretic real first cohomology of the disc, with $S$ being the
cup-product.

The standard complex structure on the unit disc
has holomorphic 1-forms that are of the form $dF$ where $F$ is a
holomorphic function on $\D$ with $F(0)=0$. Thus the boundary values of
$F$ will have only
positive index Fourier modes -- corresponding therefore
to the polarizing subspace
$W_{+}$. Now, an arbitrary point of $T(1)$ is described by the choice of
a Beltrami differential $\mu$ on $\D$ perturbing the complex structure.
We are  now asking for the holomorphic 1-forms on $\D_{\mu}$.
Solving the Beltrami equation on $\D$ provides us with
the $\mu$-conformal quasiconformal self-homeomorphism $w_{\mu}$ of the
disc.  This  $w_{\mu}$  is
a holomorphic uniformising coordinate for  the disc with the $\mu$
complex structure.  The holomorphic 1-forms subspace,
$H^{1,0}(\D_{\mu})$,
should therefore comprise those functions on $S^1$
that are the $W_{+}$ functions {\it precomposed with the boundary
values of the q.c. map $w_{\mu}$.}  That is
exactly the action of $\Pi$ on the Teichm\"uller class of $\mu$.
This explains in some detail why
$\Pi$ behaves as an infinite dimensional period mapping.

\nn
{\bf Remark:} On Segal's $C^{\Inf}$ version of the Siegel space --
constructed using Hilbert-Schmidt operators $Z$, there existed the
universal {\it Siegel symplectic metric}, which we studied in [N1] [N2] and
showed to be the same as the Kirillov-Kostant (= Weil-Petersson) metric
on $Diff(S^1)/Mob(S^1)$. For the bigger Banach manifold
$\cal S_{\infty}$  above, that pairing fails to converge on arbitrary
pairs of tangent vectors because the relevant operators are not any  more
trace-class in general. The difficulties asociated with this matter will be
addressed in Section 9 below, and in further work that is in progress.

\bigskip
\nn
{\bf 9.The universal Schottky locus and quantum calculus:}

\nn
Our object is to exhibit the image of $\Pi$ in
$\cal S_{\Inf}$. The result (equation (53)) can be recognized to be a
quantum "integrability condition" for complex structures on the circle
or the line.

\nn
{\bf PROPOSITION 9.1:}
If a positive polarizing subspace $W$ is in the ''universal Schottky
locus'', namely if $W$ is in the
image of $T(1)$ under the universal period
mapping $\Pi$, then
$W$ possesses a dense subspace which is
{\it multiplication-closed} (i.e., an ``algebra" under pointwise
multiplication modulo subtraction of mean-value.)
In quantum calculus terminology, this means that
$$
[d^{Q}_{J},J] = 0
\eqno(53)
$$
where $J$ denotes the $S$-compatible complex structure of $\Ha$ whose
$-i$-eigenspace is $W$. (Recall the various descriptions of
$\cal S_{\Inf}$ spelled out in the last section.)

\medskip
\nn
{\bf Multiplication-closed polarizing subspace:}
The notion of being multiplication-closed is well-defined
for the relevant subspaces in $\Ha_{\CC}$.
Let us note that the original polarizing subspace
$W_{+}$ contains the dense subspace of holomorphic trigonometric
polynomials (with mean zero) which constitute an algebra.
Indeed, the identity map of $S^1$ is
a member of $W_{+}$, call it $j$, and positive integral powers of $j$
clearly generate $W_{+}$ -- since polynomials in $j$ form a dense subspace
therein. Now if $W$ is any other positive polarizing subspace, we know
that it is the image of $W_{+}$ under some $T \in Sp(\Ha)$. Thus, $W$
will be multiplication-closed precisely when the image of $j$ by $T$
generates $W$, in the sense that its positive integral powers (minus the
mean values) also lie in
$W$ (and hence span a dense subspace of $W$).

In other words, we are considering $W$ ($\in \cal S_{\Inf}$ [description (a)])
to be multiplication-closed provided that the pointwise products of
functions from $W$ (minus their mean values) that happen to be $H^{1/2}$
functions actually land up in the subspace $W$ again. Multiplying $f$ and
$g$ modulo arbitrary additive constants demonstrates that this notion is
well-defined when applied to a subspace.

\nn
{\bf Quantum calculus and equation (53):} We suggest a quantum version
of complex structures in one real dimension, and note that the
integrable ones correspond to the universal Schottky locus under study.

In the spirit of algebraic geometry one takes the real Hilbert space of
functions $\Ha$ = $H^{1/2}(\RR)$ as the ``coordinate ring" of the real line.
Consequently, a complex structure on $\RR$ will be considered to be a
complex structure on this Hilbert space. Since
$\cal S_{\Inf}$ was a space of (symplectically-compatible) complex
structures on  $\Ha$, we are interpreting
$\cal S_{\Inf}$ as a space of quantum  complex structures on the line
(or circle).

Amongst the points of the universal Siegel space, those that can be
interpreted as the holomorphic function algebra for some complex
structure on the circle qualify as the ``integrable'' ones. But $T(1)$
parametrises all the quasisymmetrically related circles, and for
each one, the map $\Pi$ associates to that structure the holomorphic
function algebra corresponding to it; see the interpretation we provided
for $\Pi$ in the last section. It is clear therefore that $\Pi(T(1)$
should be the integrable complex structures. The point is that taking
the standard circle as having integrable complex structure, all the
other integrable complex structures arise from this one by a $QS$ change
of coordinates on the underlying circle. These are the complex
structures $J^{h}$ introduced in Section 6 on quantum calculus. The
$-i$-eigenspace for $J^{h}$ is interpreted as
the algebra of analytic functions on the
quantum real line with the $h$-structure. We will see in the proof
that (53) encodes just this condition.

\nn
{\bf Proof of Proposition 9.1:}  For a point of
$T(1)$ represented by a q.s. homeomorphism $\p$, the period map sends it
to the polarizing subspace $W_{\p} = W_{+} \circ \p$. But $W_{+}$ was a
multiplication-closed subspace, generated by just the identity map $j$ on
$S^1$, to start with. Clearly then, $\Pi(\p) = W_{\p}$ is also
multiplication-closed in the sense explained, and is generated by the
image of the generator of $W_{+}$ -- namely by the q.s homeomorphism
$\p$ (as a member of $\Ha_{\CC}$).
\xx

\medskip
We suspect that the converse is also true: that the $T(W_{+})$ is such
an ''algebra'' subspace for a symplectomorphism $T$ in $Sp(\Ha)$ only
when $T$ arises as pullback by a quasisymmetric homeomorphism of the circle.
This converse assertion is reminescent of standard theorems in Banach
algebras where one proves, for example, that every (conjugation-preserving)
algebra automorphism of
the algebra $C(X)$ (comprising continuous functions on a
compact Hausdorff space $X$) arises from homeomorphisms of $X$.
[Remark of Ambar  Sengupta.]
Owing to the technical hitch that $H^{1/2}$ functions are not in general
everywhere defined on the circle, we are as yet unable to find a
rigorous proof of this converse.

Here is the sketch of an idea for proving the converse.
Thus, suppose we are given a subspace $E$ that is multiplication-closed in the
sense explained. Now, $Sp(\Ha)$ acts transitively on the set of
positive polarizing subspaces. We consider a $T \in Sp(\Ha)$ that maps
$W_{+}$ to $E$ preserving the algebra structure (modulo subtracting off
mean values as usual). Denote by $j$ the identity function on $S^1$
and let $T(j)=w$ be its image in $E$.

Since $j$ is a homeomorphism and $T$ is an invertible real symplectomorphism,
one expects that $w$ is also a homeomorphism on $S^1$.  (Recall the signed
area interpretation of the canonical form (8).) It then follows
that the $T$ is nothing other that precomposition by this
$w$. That is because:
$$
T(j^{m}) = T(j)^{m} - {\rm mean ~ value} =
(w(e^{i\t}))^{m} - {\rm mean ~ value} =
j^{m} \circ w - {\rm mean ~ value.}
$$
\nn
Knowing $T$ to be so on powers of $j$ is sufficient, as polynomials
in $j$ are dense in $W_{+}$.

Again, since $T$ is the complexification of a real symplectomorphism,
seeing the action of $T$ on $W_{+}$ tells us $T$ on all of $\Ha_{\CC}$;
namely, $T$ is everywhere precomposition by that homeomorphism $w$ of
$S^1$. {\it By the necessity part of Theorem 4.1 we see that $w$ must be
quasisymmetric}, and hence that the given subspace $E$ is
the image under $\Pi$ of the Teichm\"uller point determined by $w$
(i.e., the coset of $w$ in $QS(S^1)/M\ddot{o}b(S^1)$).

\smallskip
\nn
{\bf Proof of equation (53)}: Let $J$ be {\it any} $S$-compatible complex
structure on $\Ha$, namely $J$ is an arbitrary point of $\cal S_{\Inf}$
(description (b) of Section 8). Let $J_{0}$ denote the Hilbert transform
itself, which is the reference point in the universal Siegel space;
therefore $J = T J_{0} T^{-1}$ for some symplectomorphism $T$ in
$Sp(\Ha)$. The $-i$-eigenspace for $J_{0}$ is, of course, the reference
polarizing subspace $W_{+}$, and the subspace $W$ corresponding to $J$
consists of the functions $(f+i(Jf))$ for all $f$ in $\Ha$.
Now, the pointwise product of two such typical elements of $W$ gives:
$$
(f+i(Jf))(g+i(Jg))=
[fg - (Jf)(Jg)] + i[f(Jg) + g(Jf)]
$$
\nn
In order for $W$ to be multiplication closed the function on the right
hand side must also be of the form $(h + i(Jh))$. Namely, for all relevant
$f$ and $g$ in the real Hilbert space $\Ha$ we must have:

$$
J[fg - (Jf)(Jg)] = [f(Jg) + g(Jf)]
\eqno(54)
$$

Now recall from the concepts introduced in Section 8 that one can
associate to functions $f$ their quantum derivative operators
$d^{Q}_{J}(f)$
which is the commutator of $J$ with the multiplication operator $M_{f}$
defined by $f$. The quantum derivative is being taken with respect to
any Hilbert-transform-like operator $J$ as explained above. But now a
short computation demonstrates that equation (54) is the same as saying
that:
$$
J \circ d^{Q}_{J}(f) = - d^{Q}_{J}(f)
$$
operating by $J$ on both sides shows that this is the same as (53).
That is as desired. \xx

\nn
{\bf Remark:} For the classical period mapping on the Teichm\"uller
spaces $T_g$ there is a way of understanding the Schottky locus in terms
of Jacobian theta functions satisfying the nonlinear K-P equations. In a
subsequent paper we hope to relate the finite dimensional Schottky solution
with the universal solution given above.

\nn
{\bf Remark:} For the extended period-polarizations mapping $\Pi$, the
Rauch variational formula that was exhibited in [N1], [N2], [N3],
and also here in the proof of Theorem 8.1, continues to hold.

\bigskip
\nn
{\bf 10. The Teichm\"uller space of the universal compact
lamination:}

The Universal Teichm\"uller Space, $T(1)$=$T(\Delta)$,
is a {\it non-separable} complex Banach
manifold that contains, as properly embedded complex submanifolds, all
the Teichm\"uller spaces, $T_g$, of the classical compact
Riemann surfaces of every genus $g$ ($\geq 2$). $T_g$ is $3g-3$
dimensional and appears (in multiple copies) within
$T(\Delta)$ as the Teichm\"uller space $T(G)$ of the
Fuchsian group $G$ whenever $\Delta/G$ is of genus $g$.
The closure of the union of a family of these embedded $T_g$ in
$T(\Delta)$ turns out to be a separable complex submanifold of $T(\Delta)$
(modelled on a separable complex Banach space). That submanifold
can be identified as being itself the Teichm\"uller space of the
"universal hyperbolic lamination" $H_\infty$. We will show that
$T(H_{\infty})$ carries a canonical, genus-independent version of the
Weil-Petersson metric, thus bringing back into play the K\"ahler
structure-preserving aspect of the period mapping theory.

\noindent
{\bf The universal laminated surfaces:}
Let us proceed to explain the nature of the (two possible) "universal
laminations" and the complex structures on these. Starting from any
closed topological surface, $X$,  equipped with a base point,
consider the inverse (directed) system
of all finite sheeted unbranched covering spaces of $X$ by other
closed pointed surfaces. The covering projections are all required to be
base point preserving, and isomorphic covering spaces are identified. The
{\it inverse limit space} of such an inverse system is the "lamination"
-- which is the focus of our interest.

\nn
{\it The lamination $E_{\infty}$:}
Thus, if $X$ has genus one, then, of course,
all coverings are also tori, and one
obtains as the inverse limit of the tower a certain compact
topological space -- every path component of which (the laminating
leaves) -- is identifiable with the complex plane. This space $E_{\infty}$
(to be thought of as the "universal Euclidean lamination")
is therefore a fiber space over the original torus $X$ with the fiber
being a Cantor set. The Cantor set corresponds to all the possible
backward strings in the tower with the initial element being the base
point of $X$. The total space is compact since it is a closed subset of
the product of all the compact objects appearing in the tower.

\smallskip
\nn
{\it The lamination $H_{\infty}$:}
Starting with an arbitrary $X$ of higher genus clearly produces the
{\it same} inverse limit space, denoted $H_{\infty}$, independent of the
initial genus. That is because given any two surfaces of genus greater
than one, there is always a common covering surface of higher genus.
$H_{\infty}$ is our universal hyperbolic lamination, whose Teichm\"uller
theory we will consider in this section.  For the same
reasons as in the case of $E_{\infty}$, this new lamination is also a
compact topological space fibering over the base surface $X$ with fiber
again a Cantor set. (It is easy to see that
in either case the space of backward
strings starting from any point in $X$ is an uncountable, compact,
perfect, totally-disconnected space -- hence homeomorphic to the Cantor
set.) The fibration restricted to each individual leaf (i.e., path
component of the lamination) is a universal covering projection. Indeed,
notice that the leaves of
$H_{\infty}$ (as well as of $E_{\infty}$) must  all be
simply connected -- since any non-trivial loop
on a surface can be unwrapped in a finite cover. [That corresponds to
the residual finiteness of the fundamental group of a closed surface.]
Indeed, group-theoretically speaking, covering spaces correspond to the
subgroups of the fundamental group. Utilising only normal subgroups
(namely the regular coverings) would give a cofinal inverse system and
therefore the inverse limit would still continue to be the
$H_{\infty}$ lamination. This way of interpreting things allows us to see
that the transverse Cantor-set fiber actually has a group structure.
In fact it is the pro-finite group that is the inverse limit of all the
deck-transformation groups corresponding to these normal coverings.

\medskip
\noindent
{\bf Complex structures :}
Let us concentrate on the universal hyperbolic lamination
$H_{\infty}$ from now on.
For any complex structure on $X$ there is clearly a complex
structure induced by pullback on each surface of the inverse system, and
therefore $H_{\infty}$ itself inherits a complex structure on each leaf,
so that now biholomorphically each leaf is the Poincare hyperbolic
plane. If we think of a reference complex structure on $X$, then any new
complex structure is recorded by a Beltrami coefficient on $X$, and one
obtains by pullback a complex structure on the inverse limit in the
sense that each leaf now has a complex structure and the Beltrami
coefficients vary continuously from leaf to leaf in the Cantor-set
direction. Indeed, the complex structures obtained in the above fashion
by pulling back to the inverse limit from a complex structure on any
closed surface in the inverse tower, have the special property that the
Beltrami coefficients on the leaves are locally constant in the
transverse (Cantor) direction. These "locally constant" families of
Beltrami coefficients on  $H_{\infty}$ comprise the {\it transversely
locally constant} (written ``TLC") complex structures on the lamination.
The generic complex structure on $H_{\infty}$, where all continuously
varying Beltrami coefficients in the Cantor-fiber direction are
admissible, will be a limit of the TLC subfamily of complex structures.

To be precise, a {\it complex structure} on a lamination $L$ is a
covering of $L$ by lamination charts (disc) $\times$
(transversal) so that the overlap homeomorphisms are complex
analytic on the disc direction. Two complex structures are {\it
Teichm\"uller equivalent} whenever they are related to each
other by a homeomorphism that is homotopic to the identity
through leaf-preserving continuous mappings of $L$. For us $L$
is, of course, $H_{\Inf}$. Thus we have defined the set
$T(H_{\Inf})$.

Note that there is a distinguished leaf in our lamination, namely the
path component of the point which is the string of all the base points.
Call this leaf $l$. Note that all leaves are dense in $H_{\infty}$, in
particular $l$ is dense. With respect to the base complex structure the
leaf $l$ gets a canonical identification with
the hyperbolic unit disc $\Delta$.
Hence we have the natural "restriction to $l$"  mapping of the Teichm\"uller
space of $H_{\infty}$ into the Universal Teichm\"uller space $T(l) = T(1)$.
Since the leaf is dense, the complex structure on it records the entire
complex structure of the lamination.
The above restriction map is therefore actually
{\it injective} (see [Sul]), exhibiting
$T(H_{\infty})$ as an embedded complex analytic submanifold in $T(1)$.

Indeed, as we will explain in detail below, $T(H_{\infty})$ embeds as
precisely the closure in $T(1)$ of the union
of the Teichm\"uller spaces $T(G)$ as
$G$ varies over all finite-index subgroups of a fixed cocompact Fuchsian
group. These
finite dimensional classical Teichm\"uller spaces lying within the
separable,  infinite-dimensional
$T(H_{\infty})$, comprise the TLC points of $T(H_{\infty})$.

Alternatively, one may understand the set-up at hand by looking at the
direct system of maps between Teichm\"uller spaces that is obviously
induced by our inverse system of covering maps. Indeed, each covering
map provides an immersion of the Teichm\"uller space of the covered
surface into the Teichm\"uller space
of the covering surface induced by the
standard pullback of complex structure. These immersions are
Teichm\"uller metric preserving, and provide a direct system whose
direct limit, when completed in the Teichm\"uller metric, gives
produces again $T(H_{\infty})$. The direct limit already contains the
classical Teichm\"uller spaces of closed Riemann surfaces, and the
completion corresponds to taking the closure in $T(1)$.

Let us elaborate somewhat more on these various possible
embeddings of $T(H_{\infty})$ [ which is to be thought of as the {\it
universal Teichm\"uller space of {\bf compact} Riemann surfaces}]
within the classical universal Teichm\"uller space $T(\Delta)$.

\medskip
\noindent
{\bf Explicit realizations of $T(H_\infty)$ within the universal
Teichm\"uller space:}
Start with any cocompact (say torsion-free) Fuchsian group $G$
operating on the unit disc $\Delta$,
such that the quotient is a Riemann surface $X$ of {\it arbitrary} genus $g$
greater than one.
Considering the inverse limit of the directed system of all unbranched
finite-sheeted pointed covering spaces over $X$ gives us a copy of the
universal laminated space $H_\infty$ equipped with a complex structure induced
from that on $X$. Every such choice of $G$ allows us to embed the
separable Teichm\"uller space
$T(H_\infty)$
holomorphically in the Bers universal Teichm\"uller space $T(\Delta)$.

To fix ideas, let us think of the universal Teichm\"uller space as in model
(a) of Section 2:
$ T(\Delta) = T(1) = QS(S^1)/Mob(S^1) $
For any Fuchsian group $\Gamma$ define:
$$
QS(\Gamma) = \{ w \in QS(S^1): w\Gamma w^{-1}~is~again~a~Mobius~group.\}
$$
We say that the quasisymmetric homeomorphisms in $QS(\Gamma)$ are
those that are {\it compatible} with $\Gamma$.
Then the Teichm\"uller space $T(\Gamma)$ is $QS(\Gamma)/Mob(S^1)$ clearly
sits embedded within $T(1)$. [We always think of points of $T(1)$ as
left-cosets of the form $Mob(S^1)\circ w$ = $[w]$ for arbitrary
quasisymmetric homeomorphism $w$ of the circle.]

Having fixed the cocompact Fuchsian group $G$, the Teichm\"uller space
$T(H_\infty)$
is now the closure in $T(1)$ of the direct limit of all the Teichm\"uller
spaces $T(H)$ {\it as $H$ runs over all the finite-index subgroups of the
initial cocompact Fuchsian group $G$}. Since each $T(H)$ is actually
embedded injectively within the universal Teichm\"uller space, and since
the connecting maps in the directed system are all inclusion maps, we see
that the direct limit (which is in general a quotient of the disjoint
union) in this situation is nothing other that just the set-theoretic
{\it union of all the embedded $T(H)$ as $H$ varies over all finite index
subgroups of $G$ }. This
union in $T(1)$ constitutes the dense ``TLC" (transversely locally constant)
subset of $T(H_\infty)$. Therefore, {\it the TLC subset of this embedded
copy of $T(H_\infty)$ comprises the M\"ob-classes of all those
QS-homeomorphisms that are compatible with {\it some} finite index
subgroup in $G$.}

We may call the above realization of
$T(H_\infty)$ as  {\it `` the $G$-tagged embedding" of
$T(H_\infty)$} in $T(1)$.

Remark: We see above, that just as the Teichm\"uller space of Riemann
surfaces of any genus $p$ have lots of realizations within the universal
Teichm\"uller space (corresponding to choices of reference cocompact
Fuchsian groups of genus $p$), the Teichm\"uller space of the lamination
$H_\infty$ also has many different realizations within $T(1)$.

Therefore, in the Bers embedding of $T(1)$, this realization
of $T(H_\infty)$ is the intersection of the domain $T(1)$ in the
Bers-Nehari Banach space $B(1)$ with the separable Banach subspace that
is the inductive (direct) limit of the subspaces $B(H)$ as $H$ varies
over all finite index subgroups of the Fuchsian group $G$.
(The inductive limt topology will give a complete (Banach) space; see,
e.g., Bourbaki's "Topological Vector Spaces".) It is relevant to recall that
$B(H)$ comprises the bounded holomorphic quadratic forms for the group
$H$. By Tukia's results, the Teichm\"uller space of $H$ is exactly the
intersection of the universal Teichm\"uller space with $B(H)$.

\nn
{\bf Remark:} Indeed one expects that the
various $G$-tagged embeddings of $T(H_{\Inf}$
must be sitting in general discretely separated from each other in
the Universal Teichm\"uller space. There is a result to this effect
for the various copies of $T(\Gamma)$, as the base group is varied, due to
K. Matsuzaki (preprint -- to appear in Annales Acad. Scient. Fennicae).
That should imply a similar discreteness for the family of embeddings
of $T(H_{\Inf})$ in $T(\Delta)$.

It is not hard now to see how many different copies of the Teichm\"uller
space of genus $p$ Riemann surfaces appear embedded within the
$G$-tagged embedding of $T(H_\infty)$. That corresponds to non-conjugate
(in $G$) subgroups of $G$ that are of index $(p-1)/(g-1)$ in $G$. This
last is a purely topological question regarding the fundamental group of
genus $g$ surfaces.

Modular group: One may look at those elements of the full universal
modular group $Mod(1)$ [quasisymmetric homeomorphism acting by right
translation (i.e., pre-composition) on $T(1)$] that preserve setwise the
$G$-tagged embedding of $T(H_\infty)$. Since the modular group
$Mod(\Gamma)$ on $T(\Gamma)$ is induced by right translations by those
QS-homeomorphisms that are in the normaliser of $\Gamma$:
$$
N_{qs}(\Gamma)=\{t \in QS(\Gamma): t\Gamma t^{-1}=\Gamma \}
$$
it is not hard to see that only the elements of $Mod(G)$ itself will
manage to preserve the $G$-tagged embedding of $T(H_\infty)$.
[Query: Can one envisage some limit of the modular groups
of the embedded Teichm\"uller spaces as acting on $T(H_\infty)$?]

\medskip
\noindent
{\bf The Weil-Petersson pairing:}
In [Sul], it has been shown that the tangent (and the cotangent)
space at any point of $T(H_{\Inf})$
consist of certain holomorphic quadratic
differentials on the universal lamination $H_{\Inf}$.
In fact, the Banach space $B(c)$ of tangent holomorphic quadratic
differentials at the Teichm\"uller point represented by the
complex structure $c$ on the lamination, consists
of holomorphic quadratic differentials on the leaves that vary
continuously in the transverse Cantor-fiber direction. Thus
locally, in a chart, these objects look like
$\vp(z,\lambda)dz^{2}$ in self-evident notation; ($\lambda$
represents the fiber coordinate). The lamination $H_{\Inf}$ also
comes equipped with an invariant transverse measure on the
Cantor-fibers (invariant with respect to the holonomy action of
following the leaves). Call that measure (fixed up to a scale)
$d\lambda$. [That measure appears as the limit of (normalized) measures
on the fibers above the base point that assign (at each finite
Galois covering stage) uniform weights to the points in the fiber.]
 From [Sul] we have directly therefore our present goal:

\medskip
\nn
{\bf PROPOSITION 10.1:} The Teichm\"uller space $T(H_{\Inf})$ is a
separable complex Banach manifold in $T(1)$ containing the
direct limit of the classical Teichm\"uller spaces as a dense
subset. The Weil-Petersson metrics on the classical $T_{g}$,
normalized by a factor depending on the genus, fit together and
extend to a finite Weil-Petersson inner product on
$T(H_{\Inf})$ that is defined by the formula:

$$
\int_{H_{\Inf}}^{}
\vp_{1} \vp_{2}(Poin)^{-2}dz \wedge d \oo z d\lambda
\eqno(55)
$$
where $(Poin)$ denotes the Poincare conformal factor for the Poincare
metric on the leaves (appearing as usual for all Weil-Petersson formulas).
\xx

\medskip
\nn
{\bf Remark on Mostow rigidity for $T(H_{\Inf})$:} The quasisymmetric
homeomorphism classes comprising this Teichm\"uller space are again very
non-smooth, since they appear as limits of the fractal q.s. boundary
homeomorphisms corresponding to deformations of co-compact Fuchsian
groups. Thus, the transversality proved in [NV, Part II] of the finite
dimensional Teichm\"uller spaces with the coadjoint orbit homogeneous
space $M$ continues to hold for $T(H_{\Inf})$. As explained there, that
transversality is a form of the Mostow rigidity phenomenon.
The formal Weil-Petersson
converged on $M$ and coincided with the Kirillov-Kostant metric, but
that formal metric fails to give a finite pairing on the tangent spaces
to the finite dimensional $T_{g}$. Hence there is interest in the above
Proposition.

\bigskip
\nn
{\bf 11.The Universal Period mapping and the Krichever map:}

We make some remarks on the relationship of $\Pi$ with the Krichever mapping
on a certain family of Krichever data. This could be useful in developing
infinite-dimensional theta functions that go hand-in-hand with our infinite
dimensional period matrices.

The positive polarizing subspace, $T_{\mu}(W_{+})$, that is assigned by the
period mapping $\Pi$ to a point $[\mu]$ of the universal Teichm\"uller space
has a close relationship with the Krichever subspace of $L^{2}(S^1)$ that
is determined by the Krichever map on certain Krichever data, when $[\mu]$
varies in (say) the Teichm\"uller space of a compact Riemann surface with one
puncture (distinguished point). I am grateful to Robert Penner for
discussing this matter with me.

Recall that in the Krichever mapping one takes a compact Riemann surface
$X$, a point $p \in X$, and a local holomorphic coordinate around $p$ to
start with (i.e., a member of the ``dressed moduli space''). One also
chooses a holomorphic line bundle $L$ over $X$
and a particular trivialization of $L$ over the
given ($z$) coordinate patch around $p$. We assume that the $z$ coordinate
contains the closed unit disc in the $z$-plane. To such data, the Krichever
mapping associates the subspace of $L^{2} (S^1)$ [here $S^1$ is the
unit circle in the $z$ coordinate]
comprising functions which are restrictions to that
circle of holomorphic sections of $L$ over the punctured surface
$X-\{p\}$.

If we select to work in a Teichm\"uller space $T(g,1)$ of pointed
Riemann surfaces of genus $g$, then one may choose $z$ canonically as a
certain horocyclic coordinate around the point $p$. Fix $L$ to be the canonical
line bundle $T^{*}(X)$ over $X$ (the compact Riemann surface). This has
a corresponding trivialization via ``$dz$''. The Krichever image of this
data can be considered as a subspace living on the unit horocycle
around $p$. That horocycle can be mapped over to the boundary circle
of the universal covering disc for $X-\{p\}$ by mapping out by the
natural pencil of Poincare geodesics having one endpoint at a parabolic
cusp corresponding to $p$.

{\it We may now see how to recover the Krichever subspace (for this restricted
domain of Krichever data) from the subspace in
$H^{1/2}_{\CC}(S^1)$ associated to $(X,p)$ by $\Pi$.} Recall that the
functions appearing in the $\Pi$ subspace are the boundary values of
the Dirichlet-finite harmonic functions whose derivatives give the
holomorphic Abelian differentials of the Riemann surface. Hence, to get
Krichever from $\Pi$ one takes Poisson integrals of the functions in the
$\Pi$ image, then takes their total derivative in the universal covering
disc, and restricts these to the horocycle around $p$ that is
sitting inside the universal cover
(as a circle tangent to the boundary circle of the Poincare disc).

Since Krichever data allows one to create the tau-functions of the $KP$
-hierarchy by the well-known theory of the Sato school (and the Russian
school), one may now use the tau-function from the Krichever data to
associate a tau (or theta) function to such points of our universal
Schottky locus. The search for natural theta functions associated to
points of the universal Siegel space $\cal S$, and their possible use
in clarifying the relationship between the universal and classical Schottky
problems, is a matter of interest that we are working upon.

\bigskip
\centerline {{\bf REFERENCES} }

\medskip
\item{A1.}
L.Ahlfors, Remarks on the Neumann-Poincare integral equation,
{\it Pacific J.Math}, 2(1952),  271-280.

\medskip
\item{A2.}
L.Ahlfors,  Remarks on Teichm\" uller's Space of Riemann
Surfaces, {\it Annals of Math.}, 74, 1961, 171-191.

\medskip
\item{A3.}
L. Ahlfors, {\it Lectures on quasiconformal mappings,}
Van Nostrand, (1966).

\medskip
\item{BD.}
A. Beurling and J.Deny, Dirichlet Spaces, {\it Proc. Nat. Acad. Sci.}, 45
(1959), 208-215.

\medskip
\item{C.}
A.Connes, {\it Geometrie Non Commutative}.

\medskip
\item{CS.}
A. Connes and D. Sullivan, Quantum calculus on $S^1$ and Teichm\"uller
theory, IHES preprint, 1993.

\medskip
\item{CRW.}
R.R. Coifman, R. Rochberg and G. Weiss, Factorization theorems for Hardy
spaces in several variables, {\it Annals Math.} 103 (1976),611-635.

\medskip
\item{GS.}
F. Gardiner and D. Sullivan,
Symmetric structures on a closed curve, {\it American J. Math.}
114 (1992), 683-736.

\medskip
\item{Gr.}
P. Griffiths, Periods of integrals on algebraic manifolds, {\it
Bull. American Math. Soc.,}, 75 (1970), 228-296.

\medskip
\item{HR.}
D.K.Hong and S.Rajeev,  Universal Teichm\" uller Space and
Diff$(S^{1})/S^{1}$, {\it Commun. Math. Phys.}, 135, 1991, 401-411.

\medskip
\item{KNS.}
Y.Katznelson, S.Nag and D.Sullivan,  On Conformal Welding
Homeomorphisms Associated to Jordan Curves, {\it Annales Acad. Scient. Fenn.},
A1, Math., 15, 1990, 293-306.

\medskip
\item{L.}
S. Lang, $SL_2 (\RR)$, Springer-Verlag, (1975).

\medskip
\item{Le.}
O.Lehto, {\it Univalent Functions and Teichm\" uller Spaces},
Springer Verlag, New York, 1987.

\medskip
\item{N1.}
S. Nag,
A period mapping in universal Teichm\"uller space,
{\it Bull. American Math. Soc.,} 26, (1992), 280-287.

\medskip
\item{N2.}
S. Nag,
Non-perturbative string theory and the diffeomorphism group of
the circle, in {\it ``Topological and Geometrical Methods in
Field Theory''} Turku, Finland International Symposium,
(eds. J. Mickelsson and O.Pekonen), World Scientific, (1992),
267-292.

\medskip
\item{N3.}
S. Nag,
On the tangent space to the universal Teichm\"uller space, {\it Annales
Acad. Scient. Fennicae}, vol 18, (1993), in press.

\medskip
\item{N4.}
S. Nag,
{\it The complex analytic theory of Teichm\"uller spaces,}
Wiley-Interscience, New York, (1988).

\medskip
\item{NS.}
S.Nag and D. Sullivan,
Teichm\"uller theory and the universal period mapping via
quantum calculus and the $H^{1/2}$ space on the circle,
IHES (Paris) preprint, 1st version, 1992, (submitted for publication).

\medskip
\item{NV.}
S. Nag and A. Verjovsky,
Diff $(S^1)$ and the Teichm\"uller spaces,
{\it Commun. Math. Physics}, 130 (1990), 123-138 (Part I by
S.N. and A.V. ; Part II by S.N.).

\medskip
\item{Re.}
H. Reimann, Ordinary differential equations and quasiconformal mappings,
{\it Inventiones Math.} 33 (1976), 247-270.

\medskip
\item{S.}
G. Segal,
Unitary representations of some infinite dimensional groups,
{\it Commun. in Math. Physics} 80 (1981), 301-342.

\medskip
\item{Su.}
M. Sugiura,
{\it Unitary representations and harmonic analysis - an
introduction, } 2nd ed., North Holland/Kodansha, (1975)

\medskip
\item{Sul.}
D. Sullivan, Relating the universalities of Milnor-Thurston, Feigenbaum
and Ahlfors-Bers, Milnor Festschrift volume ``Topological methods
in modern Mathematics'', (ed. L.Goldberg and A. Phillips),
Publish or Perish, 1993, 543-563.

\medskip
\item{Tr.}
H. Triebel,
{\it Theory of function spaces,}
Birkh\"auser-Verlag, (1983).

\medskip
\item{W.}
E. Witten,
Coadjoint orbits of the Virasoro group,
{\it Commun. in Math. Physics,} 114 (1981), 1-53.

\medskip
\item{Z.}
A. Zygmund,
{\it Trigonometric Series,} Vol 1 and 2,
Cambridge Univ. Press (1968).

\bigskip
\nn
\bf{The Institute of Mathematical Sciences,}

\nn
\bf{C.I.T.Campus,}

\nn
\bf{Madras  600 113, INDIA.}

\smallskip
\nn
{\it e-mail:} {\bf nag@imsc.ernet.in}

\end